\documentstyle[11pt,emulateapj,psfig]{article}

\newcommand{\etal}{et~al.\/}
\newcommand{\forbid}[3]{\mbox{\rm [{#1}\thinspace{\sc {#2}}]\thinspace{#3}\thinspace}}  
\newcommand{\drhoz}{\mbox{$\dot{\rho}_Z(z)$}}
\newcommand{\IRX}{\mbox{$F_{\rm FIR}/F_{1600}$}}
\newcommand{\Hline}[1]{\mbox{H{\footnotesize {#1}}}}
\newcommand{\Halpha}{\Hline{\mbox{$\alpha$}}}
\newcommand{\Hbeta}{\Hline{\mbox{$\beta$}}}

\newcommand{\HII}{\mbox {H\thinspace{\footnotesize II}}}
\newcommand{\irx}{\mbox{\rm IRX$_{1600}$}}
\newcommand{\Lya}{\mbox{Ly{\footnotesize$\alpha$}}}
\newcommand{\Msun}{\mbox{${\cal M}_\odot$}}

\begin{document}

\submitted{Astrophysical Journal, accepted 27 Feb., 1999}

\title{Dust Absorption And The Ultraviolet Luminosity Density At $z
 \approx 3$ \\ As Calibrated By Local Starburst Galaxies\footnotemark[1]}\footnotetext[1]{Based
on observations with the NASA/ESA {\em Hubble Space Telescope\/}
obtained at the Space Telescope Science Institute, which is operated
by the Association of Universities for Research in Astronomy, Inc.,
under NASA contract NAS5-26555.}

\author{Gerhardt R.\ Meurer, Timothy M.\ Heckman}
\affil{The Johns Hopkins University, Department of Physics and
Astronomy,\\ Baltimore, MD 21218-2686\\
Electronic mail: meurer@pha.jhu.edu,heckman@pha.jhu.edu}

\and

\author{Daniela Calzetti} 
\affil{Space Telescope Science Institute,
3700 San Martin Drive, Baltimore, MD 21218 \\ Electronic mail:
calzetti@stsci.edu}

\begin{abstract} 
We refine a technique to measure the absorption corrected ultraviolet
(UV) luminosity of starburst galaxies using rest frame UV quantities
alone, and apply it to Lyman-limit $U$-dropouts at $z \approx 3$ found
in the Hubble Deep field (HDF).  The method is based on an observed
correlation between the ratio of far infrared (FIR) to UV fluxes with
spectral slope $\beta$ (a UV color).  A simple fit to this relation
allows the UV flux absorbed by dust and reprocessed to the FIR to be
calculated, and hence the dust-free UV luminosity to be determined.
{\em International Ultraviolet Explorer\/} spectra and {\em InfraRed
Astronomical Satellite\/} fluxes of local starbursts are used to
calibrate the \IRX\ versus $\beta$ relation in terms of $A_{1600}$ (the
dust absorption at 1600\AA), and the transformation from broad band
photometric color to $\beta$.  Both calibrations are almost completely
independent of theoretical stellar population models.  We show that the
recent marginal and non-detections of HDF $U$-dropouts at radio and
sub-mm wavelengths are consistent with their assumed starburst nature,
and our calculated $A_{1600}$.  This is also true of recent observations
of the ratio of optical emission line flux to UV flux density in the
brightest $U$-dropouts.  This latter ratio turns out {\em not\/} to be a good
indicator of dust extinction.  In $U$-dropouts, absolute magnitude
$M_{1600,0}$ correlates with $\beta$: brighter galaxies are redder, as
is observed to be the case for local starburst galaxies.  This suggests
that a mass-metallicity relationship is already in place at $z \approx
3$.  The absorption-corrected UV luminosity function of $U$-dropouts
extends up to $M_{1600,0} \approx -24$ ABmag corresponding to a star
formation rate $\sim 200\, \Msun\, {\rm yr^{-1}}$ ($H_0 = 50\, {\rm km\,
s^{-1}\, Mpc^{-1}}$, $q_0 = 0.5$ assumed throughout).  The
absorption-corrected UV luminosity density at $z \approx 3$ is
$\rho_{1600,0} \geq 1.4 \times 10^{27}\, {\rm erg\, s^{-1}\, Hz^{-1}\,
Mpc^{-3}}$.  It is still a lower limit since completeness corrections
have not been done and because only galaxies with $A_{1600} \lesssim
3.6$ mag are blue enough in the UV to be selected as $U$-dropouts.  The
luminosity weighted mean dust absorption factor of our sample is $5.4
\pm 0.9$ at 1600\AA.
\end{abstract}

\keywords{galaxies: starburst -- early universe -- ultraviolet: galaxies
-- infrared: galaxies -- radio continuum: galaxies }

\section{Introduction}\label{s:intro}

\addtocounter{footnote}{1}
In the past few years we have seen the fruition of a great astronomical
endevour: the measurement of the rate of cosmological evolution via
direct censuses of star formation tracers, metal content, and metal
production rate (Lilly \etal\ 1995; Pei \&\ Fall 1995; Lilly \etal\
1996; Madau \etal\ 1996; Connolly \etal\ 1997; Madau, Pozzetti, \&\
Dickinson 1998).  This achievement has come about thanks to several key
developments including (1) deep sub-arcsecond imaging capabilities,
especially in the optical with the {\em Hubble Space Telescope\/}
(Williams \etal\ 1996), (2) multi-object spectroscopy, especially down
to $V \approx 26$ ABmag with the 10m Keck telescopes (e.g.\ Steidel
\etal 1996a,b; Lowenthal \etal\ 1997), and (3) the refinement of
photometric redshift techniques (e.g.\ Connolly \etal, 1997) especially
the Lyman-limit technique (e.g. Steidel \&\ Hamilton 1992; Steidel,
Pettini, \&\ Hamilton 1995) which allow redshift estimates to
be made to greater depth, and out to $z \approx 5$.

In a pioneering paper Madau \etal\ (1996; hereafter M96) derived the $z$
evolution of the metal production rate \drhoz\ (in comoving
coordinates), which is directly proportional to the volume density of
the star formation rate (SFR).  The ``Madau plot'' provides a direct
constraint on the evolution of galaxies, hence it's importance.  It
shows a strong (factor $\sim$ 10) decline in \drhoz\ from its peak at $z
\approx 1-2$ to the present epoch, as pointed out earlier by Lilly
\etal\ (1996).  This decline mirrors that seen in the luminosity density
of quasars (Boyle \&\ Terlevich 1997).  On the other side of the peak,
at $z \approx 4$, \drhoz\ is apparently only equal to that of the
present epoch.  Subsequent work by Connolly \etal\ (1997) fills in the
$z = 1 -2$ gap and is consistent with M96.  The overall shape of the
curve is consistent with models of the observed evolution in the
properties of damped Lyman-$\alpha$ systems (Pei \&\ Fall, 1995).  This
form for \drhoz\ is consistent with hierarchical models of galaxy
formation (e.g.\ Searle \&\ Zinn, 1978; Zepf \&\ Ashman, 1993; Baugh
\etal\ 1998).

One concern about the Madau plot is that it is heavily biased to
observations in the rest frame UV.  The problem is that dust efficiently
scatters and absorbs UV radiation, greatly complicating the
interpretation of the detected UV emission.  Not only is the amount of
dust important, but so is the geometry (where it is relative to the
stars), and the composition of the dust.  For example, in the Local
Group the UV ``extinction law'' varies from galaxy to galaxy and within
galaxies, (e.g.\ Fitzpatrick 1986).  The term ``extinction law'' was
originally defined for stars.  It quantifies the wavelength dependent
combined effect of dust absorption and scattering out of the line of
sight toward point sources.  The geometry is quite different in distant
galaxies where individual stars can not be isolated, and instead
galaxies in their entirety, or large portions of them are observed.  In
such cases much of the scattered light may be recovered (by scattering
into the line of sight), and it matters a great deal how porous the dust
distribution is and whether the stars are mixed with the dust or not
(e.g.\ Witt, Thronson, \&\ Capuano 1992; Calzetti, Kinney \&\
Storchi-Bergmann 1994).  There are a variety of terms for the wavelength
dependent net dimming of the light including ``effective extinction
law'' (Calzetti \etal\ 1994) and ``obscuration law'' (Calzetti 1997).
Here we adopt the term ``absorption law'' to emphasize that it is dust
absorbing photons that causes the dimming.  Absorption laws can be
derived empirically (e.g.\ Calzetti \etal\ 1994; Calzetti 1997) and
hence average together all the above geometric factors into a simple
wavelength dependent correction.

Observations of high redshift galaxies in the sub-mm and mm bands prove
that molecular gas and dust formation have occurred as far back as $z =
4.7$ (Omont \etal\ 1996; Ivison \etal\ 1998; Guilloteau \etal\ 1997;
Frayer \etal\ 1999).  Especially interesting are the recent spate of
blank field sub-mm detection with SCUBA, which we discuss in more detail
in \S\ref{ss:rc} and \S\ref{ss:scuba}.  These can account for a
significant fraction of the sub-mm background, and are mostly attributed
to dusty star forming galaxies having $z > 1$.  Following from UV
observations of local starbursts (Meurer \etal\ 1995, hereafter Paper I)
we have argued that the correction factor to \drhoz\ is
considerable, a factor $\sim 15$ at $z \approx 3$ (Meurer \etal, 1997;
hereafter Paper II), while others (e.g.\ Madau 1997; Pettini \etal\
1997) prefer fairly modest, $\sim$ factor of three, corrections.  The
amount of high-$z$ absorption has a direct bearing on the interpretation
of the Madau plot; large corrections at high-$z$ can push back the peak
$\dot{\rho}_Z$, as might be expected by monolithic collapse dominated
galaxy formation (e.g. Eggen, Lynden-Bell, \&\ Sandage 1962; Bower,
Lucey \&\ Ellis 1992; Ortolani \etal\ 1995).

Here we reexamine the UV luminosity density at high-$z$, with particular
emphasis on the $U$-dropouts in the HDF.  Our technique has two
strengths.  Firstly, it is based on the similarity between local
starburst galaxies and Lyman limit systems as summarized in
\S~\ref{s:comp}.  The main assumption we make is that the high redshift
galaxies have the same multi-wavelength spectral properties as local
starbursts.  We use local starburst galaxies to calibrate the
relationship between UV reddening and UV absorption (in \S~\ref{s:cal})
thus largely avoiding the uncertainties in population model fitting and
the choice of effective UV absorption laws.  Secondly, our calibration
of the UV absorption law recovers the intrinsic (i.e. dust-free) UV
emission that is absorbed by dust and reradiated in the far infrared.
Hence, up to the modest amounts of absorption considered here (a few
magnitudes), UV observations are capable of recovering the total
bolometric luminosity of starbursts.  We apply our calibrations to the
$U$-dropouts in the HDF in \S~\ref{s:appl}.  In \S~\ref{s:test} we
demonstrate the validity of our starburst assumption by showing that new
constraints on the SEDs of these high-$z$ systems are consistent with
the SEDs of local UV selected starbursts.  Implications of our results
are discussed in \S\ref{s:disc}, while \S~\ref{s:sum} summarizes the
paper.

Throughout this paper we adopt $H_0 = 50\, {\rm km\, s^{-1}\,
Mpc^{-1}}$, $q_0 = 0.5$.

\section{High-$z$ galaxies and local starbursts}\label{s:comp}

Progress in understanding the Lyman-dropout galaxies would be
accelerated if we could find analogs to these galaxies in the local
universe. Fortunately, there is mounting evidence that the Lyman dropout
galaxies indeed resemble starburst galaxies in many respects.

While the magnitude of the extinction corrections to the rest-frame UV
light (the subject of the present paper) is still a matter of debate, it
is nevertheless clear that local starbursts have surface-brightnesses
that are similar to the distant galaxies (Paper II).  That is, the
high-z galaxies appear to be `scaled-up' versions of the local
starbursts with the same SFRs per unit area and
comparable stellar surface-mass-densities. In this regard, starbursts
are a far better match to the Lyman-dropout galaxies than are ordinary
late-type galaxies, which have SFRs per unit area that
are several orders-of-magnitude smaller (cf.\ Paper II; Kennicutt 1998).

The overall rest-frame UV-through-optical spectral properties of local
starbursts and the Lyman dropout galaxies are similar in terms of the
spectral energy distribution (cf. Sawicki \& Yee 1998) and the dominance
of the UV and optical spectra by absorption and emission lines
respectively (cf.\ Leitherer 1997; Pettini \etal\ 1998). The Lyman limit
systems in the HDF are clearly spatially resolved (Giavalisco, Steidel,
\&\ Macchetto 1996), demonstrating that their rest frame UV light is not
dominated by an AGN.  In addition the spectra of the Lyman-dropout
galaxies is more similar to starburst galaxies than AGN.  Indeed the
typical strengths of the UV stellar and interstellar absorption-line
features are comparable in starbursts and Lyman-dropout galaxies,
indicative of broadly similar stellar and interstellar content
(e.g. Conti, Leitherer, \&\ Vacca 1996; Lowenthal \etal\ 1997).  The
local starbursts and Lyman dropout galaxies also have similar
distributions of observed UV colors, (Paper II, and here in
\S\ref{s:appl}), - which we argue chiefly measures the amount of
dust-reddening.

Even the dynamical state of the interstellar medium in the Lyman
dropout galaxies resembles that in local starbursts. Both are
driving extensive outflows of gas at velocities of several hundred
km s$^{-1}$. The signature of this is the systematic blueshift of
the broad UV interstellar absorption-lines with respect to the
rest-frame optical (e.g. H$\beta$) and/or rest-frame UV
(Ly$\alpha$) nebular emission-lines (Pettini \etal\ 1998; Franx
\etal\ 1997; Kunth \etal\ 1998; Gonzalez-Delgado \etal\ 1998). The
velocity dispersion of the emission-line gas is also similar in the
two galaxy classes (Pettini \etal\ 1998).

To reiterate, in all their properties measured to-date, the Lyman
dropout galaxies appear to be simply scaled-up versions of local
starbursts: larger and more luminous, but otherwise remarkably similar.
The challenge is to determine the intrinsic luminosities from the UV
fluxes and colors of dusty Lyman limit systems.  The above similarities
justifies our approach of calibrating the effects of dust on local
starbursts and applying it to the high-$z$ galaxies.

\section{The local sample calibration}\label{s:cal}

\begin{figure*}
\centerline{\hbox{\psfig{figure=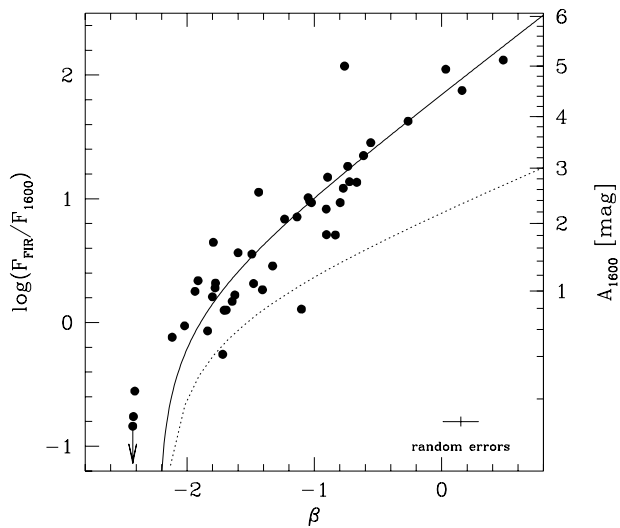,width=8.5cm}}}
\caption{The ratio of FIR to UV flux at 1600\AA\ compared to UV
spectral slope $\beta$ for UV selected starburst galaxies.  The right
axis converts the flux ratio to 1600\AA\ absorption $A_{1600}$ using
eq.~\ref{e:afit}.  The solid line shows our adopted linear fit to the
$A_{1600}$, $\beta$ relationship.  The dotted line shows the proposed
dust absorption/population model of Pettini \etal\ (1997).
\label{f:uvir}}
\end{figure*}

Figure~\ref{f:uvir} illustrates the heart of the technique.  It shows
the relationship between the ratio of far infrared (FIR) and ultraviolet
(UV) fluxes (here at 1600\AA) and the UV spectral slope $\beta$
(equivalent to a color) for a sample of starburst galaxies.  Since the
FIR flux is due to dust radiatively heated by the absorbed UV radiation,
the $y$ axis is a measurement of dust absorption.  This figure then
shows that dust absorption is correlated with UV reddening.  Such a
relationship is expected for dust predominantly located in a foreground
screen (Witt \etal\ 1992; Paper I), although this screen need not be
homogeneous (Calzetti \etal\ 1994; Calzetti 1997).  Because this
correlation links $F_{\rm FIR}$ to UV quantities, then regardless of the
exact dust geometry, this figure provides a powerful {\em empirical\/}
tool for recovering the radiation reprocessed by dust and thus
determining the total absorption-corrected UV flux of starbursts, using
UV quantities alone.

In the next sub-section we define our local calibrating samples and the
quatities we use.  In order to apply this tool and interpret the results
we perform three calibrations.  1.\ \IRX\ is calibrated in terms of
absorption at 1600\AA\ and this is fitted as a function of $\beta$.  2.\
The spectroscopic index $\beta$ is calibrated in terms of broad-band
colors (with a $z$ correction), since photometric colors are easier to
measure than spectroscopic ones.  3.\ The relationship between
luminosity measured at 1600\AA\ is related to SFR.
These calibrations are detailed in the final three sub-sections.

\subsection{Sample and definitions}

The local sample used to derive the various calibrations in this paper
is listed in Table~\ref{t:loc}.  It includes the sources shown in
Fig.~\ref{f:uvir}, and is drawn from the {\em International
Ultraviolet Explorer\/} (IUE) atlas of Kinney \etal\ (1993), and
contains galaxies having an ``activity class'' consistent with being a
starburst, i.e.\ SB nuc.\ (starburst nucleus), SB ring (starburst ring),
BCDG (blue compact dwarf galaxy), or BCG (blue compact galaxy).
Although the UV sources in these galaxies tend to be centrally
concentrated we find that galaxies with optical diameters $D_{25} > 4'$
($D_{25}$ is measured at $B = 25$ mag arcsec$^{-2}$) tend to fall above
the relationship shown in Fig.~\ref{f:uvir}.  This is probably because
significant UV emission extends beyond the $20'' \times 10''$ IUE
aperture.  Hence these large galaxies were excluded from the
sample\footnote{The other galaxies that were excluded are NGC~1569 because
of excessive foreground Galactic extinction; NGC~3690 because the IUE
pointing is likely to be wrong (Paper I); the BCDGs Mrk209, Mrk220,
and Mrk499 because they have neither IRAS fluxes nor upper limits; and
finally five galaxies with low S/N IUE spectra in the Kinney \etal\
atlas: NGC~4853, IC~2184, Mrk~309, Mrk~789, UGC~6448.}.
Further limiting the sample to galaxies having $D_{25} < 2.5'$ removes a
few more outliers, but not just galaxies above the relationship.
Furthermore, it also severely depletes the points with $\beta >
-0.5$. Applying fits to data limited in this way changes our final
$\rho_{1600}(z=2.75)$ results by $< 10$\%, hence we retain $D_{25} < 4'$
as the diameter limit for the local sample.  The lack of systematic
residuals for galaxies up to 12 times larger than the IUE aperture
indicates that the UV emission of these galaxies is very compact, e.g.\
in a circum-nuclear starburst.

{\scriptsize
\begin{deluxetable}{l r r r r r c c l}
\tablecaption{Local sample: relevant data\label{t:loc}}
\tablehead{\colhead{name} & \colhead{$A_B({\rm Gal})$} & \colhead{$\beta$} & 
\colhead{$\log(\frac{F_{\rm FIR}}{F_{1600}})$} & \colhead{$\log(F_{\rm FIR})$} & 
\colhead{$\log(\frac{F_{\rm H\beta}}{f_{1600}})$} & 
\colhead{$\frac{F_{\rm H\alpha}}{F_{\rm H\beta}}$} &
\colhead{$\frac{F_{\rm [OIII]5007}}{F_{\rm H\beta}}$ } & \colhead{Relevant} \nl
\colhead{} & \colhead{\small [mag]} & \colhead{} & \colhead{} & \colhead{\small [${\rm erg/cm^2/s}$]} & 
\colhead{\small [\AA]} & \colhead{} & \colhead{} & \colhead{figures}}
\startdata
NGC~4861     & 0.00 & $-2.46$ & $ -0.08$ & $ -9.97$ & \nodata & \nodata & \nodata & 3  \nl
I Zw 18      & 0.02 & $-2.43$ & $< -0.84$ & $< -11.50$ & \nodata & \nodata & \nodata & 1,3 \nl
NGC~1705     & 0.18 & $-2.42$ & $ -0.76$ & $-10.29$ & $-0.01$ &   2.85  &   4.61  & 1,7\nl
Mrk~153      & 0.00 & $-2.41$ & $ -0.55$ & $-10.92$ & \nodata & \nodata & \nodata & 1 \nl
Tol1924--416 & 0.30 & $-2.12$ & $ -0.12$ & $-10.17$ & $ 0.94$ &   2.91  &   4.87  & 1,3,7 \nl
UGC~9560     & 0.00 & $-2.02$ & $ -0.03$ & $-10.41$ & $ 0.79$ &   3.36  &   4.13  & 1,3,7 \nl
Mrk~66       & 0.01 & $-1.94$ & $  0.25$ & $-10.56$ & $ 0.68$ &   2.85  &   2.90  & 1,7 \nl
NGC~3991     & 0.02 & $-1.91$ & $  0.34$ & $ -9.80$ & \nodata & \nodata & \nodata & 1,3 \nl
NGC~3738     & 0.00 & $-1.89$ &  \nodata &  \nodata & \nodata & \nodata & \nodata & 3 \nl
UGCA~410     & 0.02 & $-1.84$ & $ -0.07$ & $-10.86$ & \nodata & \nodata & \nodata & 1 \nl
Mrk~357      & 0.17 & $-1.80$ & $  0.21$ & $-10.37$ & $ 0.80$ &   3.25  & \nodata & 1,7 \nl
NGC~3353     & 0.00 & $-1.79$ & $  0.65$ & $ -9.61$ & \nodata & \nodata & \nodata & 1,3 \nl %
MRK~54       & 0.02 & $-1.78$ & $  0.28$ & $-10.27$ & \nodata & \nodata & \nodata & 1 \nl
NGC~1140     & 0.11 & $-1.78$ & $  0.32$ & $ -9.77$ & $ 0.96$ &   3.18  &   2.66  & 1,7 \nl
Mrk~36       & 0.00 & $-1.72$ & $ -0.26$ & $-10.94$ & \nodata & \nodata & \nodata & 1 \nl
MCG6-28-44   & 0.02 & $-1.77$ & $  0.31$ & $-10.27$ & \nodata & \nodata & \nodata & 1 \nl
NGC~1510     & 0.00 & $-1.71$ & $  0.10$ & $-10.36$ & $ 0.67$ &   3.11  &   4.95  & 1,3,7 \nl
MRK~19       & 0.14 & $-1.69$ & $  0.10$ & $-10.64$ & \nodata & \nodata & \nodata & 1 \nl
NGC~4214     & 0.00 & $-1.69$ &  \nodata &  \nodata & \nodata & \nodata & \nodata & 3 \nl
NGC~4670     & 0.04 & $-1.65$ & $  0.17$ & $ -9.85$ & \nodata & \nodata & \nodata & 1,3 \nl
NGC~1800     & 0.00 & $-1.65$ &  \nodata &  \nodata & $ 0.41$ &   3.08  &   3.99  & 7 \nl
UGC~5720     & 0.00 & $-1.62$ & $  0.22$ & $ -9.97$ & \nodata & \nodata & \nodata & 1,3 \nl
UGC~5408     & 0.00 & $-1.60$ & $  0.56$ & $-10.22$ & \nodata & \nodata & \nodata & 1 \nl
NGC~7673     & 0.14 & $-1.50$ &  \nodata &  \nodata & $ 0.75$ &   4.46  &   2.13  & 3,7 \nl
NGC~3125     & 0.25 & $-1.49$ & $  0.55$ & $ -9.64$ & $ 1.07$ &   3.28  &   4.72  & 1,7 \nl %
Haro 15      & 0.08 & $-1.48$ & $  0.31$ & $-10.16$ & $ 0.74$ &   2.85  &   2.32  & 1,7 \nl
NGC2537      & 0.15 & $-1.44$ & $  1.05$ & $ -9.73$ & \nodata & \nodata & \nodata & 1 \nl
UGC~3838     & 0.07 & $-1.41$ & $  0.26$ & $-10.55$ & \nodata & \nodata & \nodata & 1 \nl
NGC~7793     & 0.00 & $-1.34$ &  \nodata &  \nodata & $ 0.63$ &   4.60  &   0.55  & 3,7 \nl
NGC~5253     & 0.18 & $-1.33$ & $  0.69$ & $ -8.86$ & $ 1.02$ &   3.18  &   4.68  & 3,7 \nl
NGC~7250     & 0.62 & $-1.33$ & $  0.46$ & $ -9.77$ & $ 0.80$ &   3.14  &   3.48  & 1,7 \nl
Mrk~542      & 0.07 & $-1.32$ &  \nodata &  \nodata & $ 0.54$ &   4.00  &   0.79  & 7 \nl
NGC~7714     & 0.15 & $-1.23$ & $  0.84$ & $ -9.32$ & $ 1.18$ &   4.36  &   1.60  & 1,3,7 \nl
Mrk~487      & 0.02 & $-1.15$ &  \nodata &  \nodata & $ 1.01$ &   3.25  &   6.61  & 7 \nl
NGC~3049     & 0.04 & $-1.14$ & $  0.85$ & $ -9.84$ & $ 0.99$ &   4.00  &   0.23  & 1,7 \nl
UGC~6456     & 0.09 & $-1.10$ & $  0.11$ & $-10.74$ & \nodata & \nodata & \nodata & 1 \nl
NGC~3310     & 0.00 & $-1.05$ & $  1.01$ & $ -8.83$ & \nodata & \nodata & \nodata & 1 \nl
NGC~5996     & 0.06 & $-1.04$ & $  0.98$ & $ -9.64$ & $ 0.86$ &   4.76  &   0.38  & 1,7 \nl %
NGC~4385     & 0.03 & $-1.02$ & $  0.97$ & $ -9.65$ & $ 0.97$ &   5.48  &   0.79  & 1,7 \nl
Mrk~499      & 0.00 & $-1.02$ &  \nodata &  \nodata & $ 0.38$ &   4.65  &   3.57  & 7 \nl
NGC~5860     & 0.03 & $-0.91$ & $  0.92$ & $-10.04$ & $ 0.88$ &   6.04  &   0.64  & 1,7 \nl
IC~1586      & 0.07 & $-0.91$ & $  0.71$ & $-10.28$ & $ 0.89$ &   5.30  &   1.35  & 7 \nl
NGC~2782     & 0.00 & $-0.90$ & $  1.17$ & $ -9.33$ & \nodata & \nodata & \nodata & 1 \nl
ESO383-44    & 0.13 & $-0.83$ & $  0.71$ & $-10.13$ & \nodata & \nodata & \nodata & 1 \nl
NGC~5236     & 0.14 & $-0.83$ &  \nodata &  \nodata & $ 0.69$ &   3.91  &   0.22  & 7 \nl
NGC~2415     & 0.16 & $-0.80$ & $  0.97$ & $ -9.36$ & \nodata & \nodata & \nodata & 1 \nl
MCG--01--30--33 & 0.05 & $-0.77$ & $  1.09$ & $ -9.84$ & \nodata & \nodata & \nodata & 1 \nl 
NGC~1614     & 0.22 & $-0.76$ & $  2.07$ & $ -8.84$ & $ 1.24$ &   7.85  &   0.96  & 7 \nl
NGC~6217     & 0.14 & $-0.74$ & $  1.26$ & $ -9.22$ & $ 0.79$ &   5.08  &   0.39  & 1,7 \nl
NGC~6052     & 0.10 & $-0.72$ & $  1.14$ & $ -9.48$ & $ 1.04$ &   3.54  &   2.01  & 1,7 \nl
NGC~4500     & 0.00 & $-0.67$ & $  1.13$ & $ -9.69$ & \nodata & \nodata & \nodata & 1 \nl
IC~214       & 0.11 & $-0.61$ & $  1.35$ & $ -9.56$ & $ 0.66$ &   5.08  &   1.82  & 7 \nl
NGC~3504     & 0.02 & $-0.56$ & $  1.45$ & $ -8.96$ & \nodata & \nodata & \nodata & 1 \nl %
NGC~4194     & 0.00 & $-0.26$ & $  1.63$ & $ -8.99$ & $ 1.24$ &   6.81  &   1.13  & 1,7 \nl
NGC~2798     & 0.01 & $ 0.03$ & $  2.05$ & $ -9.01$ & \nodata & \nodata & \nodata & 1 \nl
NGC~3256     & 0.55 & $ 0.16$ & $  1.88$ & $ -8.36$ & $ 1.23$ &   5.48  &   0.47  & 1,7 \nl
NGC~7552     & 0.00 & $ 0.48$ & $  2.12$ & $ -8.44$ & $ 1.29$ &   6.11  &    0.1  & 1,3,7 \nl
\enddata
\end{deluxetable} }

The IUE spectra in the Kinney \etal\ atlas were measured to determine
the UV quantities required for Fig.~\ref{f:uvir}.  The ultraviolet flux
at 1600\AA, $F_{1600}$ is a generalized flux of the form $F_\lambda =
\lambda f_\lambda$, and $f_\lambda$ is the flux density per wavelength
interval.  It was measured with the {\sl IRAF / STSDAS}\footnote{The
{\sl Image Reduction and Analysis Facility} software package is
distributed by the National Optical Astronomy Observatories, which are
operated by the Association of Universities for Research in Astronomy,
Inc., under cooperative agreement with the National Science
Foundation. {\sl STSDAS\/} is the {\sl Space Telescope Science Data
Analysis Software\/} package for IRAF, distributed by STScI.} package
{\sl SYNPHOT\/} employing a square passband with a central wavelength of
1600\AA\ and width of 350\AA.  This filter is meant to approximate the
rest frame parameters of the standard WFPC2 filters F606W and F814W for
objects with redshifts $z = 2.75$, and 4, respectively (i.e.\ $U$ and
$B$ band dropouts).  The ultraviolet spectral slope $\beta$ is
determined from a power law fit of the form
\begin{equation}
f_\lambda \propto \lambda^\beta.
\end{equation}
to the UV continuum spectrum as defined by the ten (rest wavelength)
continuum bands listed by Calzetti \etal\ (1994).  These spectral fits
were performed after first removing Galactic extinction using the law of
Seaton (1979) and taking Galactic extinction values $A_B = 4.1E(B-V)$
from Burstein \&\ Heiles (1982, 1984) as listed by NED\footnote{The
NASA/IPAC Extragalactic Database (NED) is operated by the Jet Propulsion
Laboratory, California Institute of Technology, under contract with the
National Aeronautics and Space Administration. }.  Since the continuum
spectrum is never exactly a pure power law, $\beta$ is subject to
systematic uncertainties due to the placement of the continuum windows.
Eighteen of the data points in Fig~\ref{f:uvir} represent galaxies observed
with one of IUE's short wavelength (SW) cameras ($\lambda \approx
1100$\AA\ to 1975\AA) only and not with either of the LW cameras
($\lambda \approx 1975$\AA\ to 3000\AA), hence they do not have data in
Calzetti's \etal\ (1994) tenth window ($\lambda = 2400 - 2580$\AA).
Their $\beta$ values were determined from the SW only measurements using
the following relationship:
\begin{equation}
\langle \beta - \beta_{\rm SW} \rangle = -0.16 \pm 0.04.\label{e:swoff}
\end{equation}
This was determined from measuring $\beta$ of 16 high signal-to-noise
IUE spectra with no noticeable SW/LW break, both with and without the
tenth window.  The uncertainty is the standard error on the mean.

The only non-UV quantity in Fig.~\ref{f:uvir} is the far-infrared flux
$F_{\rm FIR}$ which is derived from {\em InfraRed Astronomical
Satellite\/} (IRAS)  60 and 100\micron\ flux
densities listed by NED and is defined by Helou \etal\ (1988).  For
convenience in notation we define the ratio $\IRX = \irx$

\subsection{Errors\label{ss:err}}

Typical error bars are shown in the bottom right of fig~\ref{f:uvir}.
We estimate an uncertainty in $\beta$ of 0.14 from the $\chi^2$ fits to
the spectra, and the scatter about the mean relationship given in
eq.~\ref{e:swoff}).  We adopt an average uncertainty in $F_{1600}$ of
5\%, which we showed in Paper I is the typical level of agreement
between broad band HST UV fluxes and IUE spectra.  This is higher than
expected from the error spectra presented with the Kinney \etal\ (1993)
IUE atlas, which are commonly at the 10\%\ -- 20\%\ level {\em per
pixel\/}, and hence $<1$\% when integrated over the UV passband.  The
level of agreement between HST and IUE is worse probably because of the
centering errors with IUE.  The median uncertainty in $F_{\rm FIR}$ for
our sample is 7\%, derived from the uncertainties in $F_{60}$ and
$F_{100}$ as listed by NED.  Combining the $F_{1600}$ and $F_{\rm FIR}$
errors we estimate that the error in $\log(\IRX)$ is 0.04 dex.

These uncertainties represent random errors.  Systematic errors will
effectively move all data points with respect to the zeropoints of the
axes.  C.\ Robert (1999, private communication) estimates that there is
a sytematic uncertainty of 0.15 in $\beta$ due to the placement of the
continuum windows in the spectrum fit.  

Figure~\ref{f:uvir} illustrates that the random errors in the quantities
are smaller than the scatter about the relationship.  Hence, the scatter
in Fig.~\ref{f:uvir} is largely intrinsic.  Possible sources of
intrinsic scatter are dust geometry, dust composition, and intrinsic
spectral slope (which depends on the star formation history), all of
which may vary from galaxy to galaxy.

\subsection{$\beta$ -- $A_{1600}$ calibration\label{ss:bacal}}

We assume that $F_{\rm FIR}$ is due to the thermal emission of dust
heated by the radiation it absorbs.  This consists primarily of
non-ionizing photons with the addition of \Lya\ photons which are
resonantly scattered by Hydrogen until they are absorbed by dust grains.
We assume that ionizing photons do not significantly contribute to dust
heating\footnote{Starburst population models indicate that typically
only $\leq 10$\%\ of the bolometric luminosity of starbursts is emitted
below 912\AA\ (Leitherer \&\ Heckman, 1995).  The fraction of this that
is directly absorbed by dust grains is estimated to be $\sim 25$\%
(Smith, Biermann, \&\ Mezger 1978).}.  For these assumptions \irx\ can be
written as
{\scriptsize
\begin{eqnarray} 
\irx\! & \!\!\equiv\! & \!\!\frac{F_{\rm FIR}}{F_{1600}} \nonumber \\
    \! & \!\!=\! & \!\!\left( \frac{F_{\rm Ly\alpha} + 
\int_{\rm 912\AA}^\infty f_{\lambda',0}(1 - 10^{-0.4A_{\lambda'}})d\lambda'}
{F_{1600,0}10^{-0.4A_{1600}}} \right)\!\!
\left( \frac{F_{\rm FIR}}{F_{\rm bol}}\right)_{\rm Dust}\label{e:irx1}
\end{eqnarray} }
$f_{\lambda,0}$ is the unattenuated flux density of the emitted
spectrum, $A_\lambda$ is the net absorption in magnitudes by dust as a
function of wavelength, and $F_{\rm Ly\alpha}$ is the \Lya\ flux.  In
our models $F_{\rm Ly\alpha}$ is derived from the spectrum shortward of
912\AA\ using the standard assumption that each ionizing photon results
in one recombination and consequent decay down to case-B population
levels.  Customarily one adopts a form of the absorption (or extinction)
law scaled by the optical reddening, e.g.\ $A_\lambda = X_\lambda
E(B-V)$.  In eq.~\ref{e:irx1} the first term gives the bolometric flux
of the absorbed radiation, normalized by $F_{1600}$, and the second term
gives the fraction of the bolometric flux emitted by dust that is
intercepted by the FIR passband (e.g.\ Helou \etal\ 1988).

Strictly speaking \irx\ in our model depends on the intrinsic
spectrum and the absorption curve, even at wavelengths well outside the
1600\AA\ passband.  However, for dust heating dominated by young stellar
populations eq.~\ref{e:irx1} is well approximated by the formula
{\scriptsize
\begin{equation}
\irx = \left(10^{0.4A_{1600}}-1\right)
\left(\frac{F_{\rm Ly\alpha}+\int_{\rm 912\AA}^\infty f_{\lambda',0}d\lambda'}
{F_{1600,0}} \right)
\left( \frac{F_{\rm FIR}}{F_{\rm bol}}\right)_{\rm Dust}.\label{e:irx2}
\end{equation} }
Here the first term gives the fraction of $F_{1600,0}$ that is absorbed
by dust; the second term is similar to a bolometric correction, giving
the maximum amount of heating available to the dust divided by the
intrinsic 1600\AA\ flux; and the third term is the same as the second
term in eq.~\ref{e:irx1}.  The ratio of eqs.~\ref{e:irx1} and
\ref{e:irx2} asymptotes to unity as $A_\lambda \rightarrow \infty$ for
all $\lambda$.  We have calculated this ratio under a variety of
assumptions applicable to starbursts.  The constant star formation rate
(CSFR) solar metallicity models of Leitherer and Heckman (1995) having a
Salpeter (1955) IMF slope and durations of $\Delta t = 1$ to 100 Myr
were used as the initial spectra.  Leitherer and Heckman (1995) varied
other parameters in their population models including metallicity, IMF
slope and cutoff, and other star formation histories (i.e. instantaneous
bursts).  They find that as long as the population is ionizing (i.e.\
produces a strong ionizing flux, as expected for starbursts by their
strong emission line spectra), these parameters have very little effect
on the overall UV spectrum which has $\beta_0 = -2.0$ to $-2.6$ in all
cases.  This range in intrinsic $\beta$ is well covered by the models we
examine here.  The absorption was parameterized by $E(B-V) = 0.01 $ to 1
and a variety of forms of $X_\lambda$ including the following extinction
laws (assuming extinction = absorption): Galactic (Seaton, 1979), LMC
(Howarth, 1983), and SMC (using a polynomial fit to the Bouchet \etal\
1985 curve); and the starburst absorption laws of Kinney \etal\ (1994),
Calzetti \etal\ (1994), and Calzetti \etal\ (1997).  In all cases the
ratio of equations \ref{e:irx1} and \ref{e:irx2} is within 30\%\ of
unity for $A_{1600} \gtrsim 0.3$ mag.  We conclude that eq.~\ref{e:irx2}
is a good approximation of \irx\ when the dust is heated by
intrinsically young stellar populations and as long as the form of
$X_\lambda$ is not extremely different from standard
absorption/extinction laws.

Equation~\ref{e:irx2} can be rewritten as 
\begin{equation}
\irx = (10^{0.4A_{1600}} - 1)B,
\label{e:irxlin}
\end{equation}
where B is the ratio of the two bolometric like corrections:
\begin{equation}
B = \frac{\rm BC(1600)_\star}{\rm BC(FIR)_{Dust}},
\end{equation}
\begin{equation}
{\rm BC(1600)_\star} = \left( \frac{F_{\rm Ly\alpha}
+\int_{\rm 912\AA}^\infty f_{\lambda',0}d\lambda'}{F_{1600,0}}\right)_\star
\label{e:bcstar}
\end{equation}
\begin{equation}
{\rm BC(FIR)_{dust}} = \left( \frac{F_{\rm bol}}{F_{\rm FIR}}\right)_{\rm Dust}
\end{equation}
Here we assume that $B$ is constant, and estimate the best value for it.
We do this by cross comparing observations with models of the
photometric properties of stellar populations (${\rm BC(1600)_\star}$)
and dust emission (${\rm BC(FIR)_{dust}}$).

Values of ${\rm BC(1600)_\star}$ were calculated for the CSFR population
models of Leitherer \&\ Heckman (1995) with standard Salpeter (1955) IMF
slope, and an upper mass limit of 100 \Msun.  For burst durations in the
range of 0 to 100 Myr, we find $1.56 \leq {\rm BC(1600)_\star} \leq
1.76$.  We also calculated ${\rm BC(1600)_\star}$ directly from the
observed SED of NGC~1705, one of the least extincted sources in our local
sample.  We started with the observed IUE + optical SED published by
Storchi-Bergmann \etal\ (1995) after correction for Galactic extinction.
This was extended down to $\lambda = 912$\AA\ by extrapolation using the
power law fit (yielding $\beta = -2.40$).  The SED was extended redwards
to $\lambda = 2.4\, \mu$m using the {\it JHK} photometry of Jones as
reported in Meurer \etal\ (1992)\footnote{Their {\sl I} band photometry
was used to calculate a 0.34 mag aperture correction to match the IUE
size aperture of Storchi-Bergmann \etal\ (1995).}.  The total \Halpha\
flux given by Meurer \etal\ (1992) was converted to $F_{\rm Ly\alpha}$
assuming case B recombination. The integral in eq.~\ref{e:bcstar} was
evaluated directly from the SED, and $F_{1600,0}$ was measured with
SYNPHOT.  This yields ${\rm BC(1600)_\star} = 1.81$.  Note that this may
be a bit overestimated (by up to 20\%) since there appears to be a break
between the SW and LW portions of the IUE spectrum, with the SW portion
being relatively underestimated.  We conclude that within the
uncertainties the NGC~1705 measurement is consistent with the model
estimates.

\begin{figure*}
\centerline{\hbox{\psfig{figure=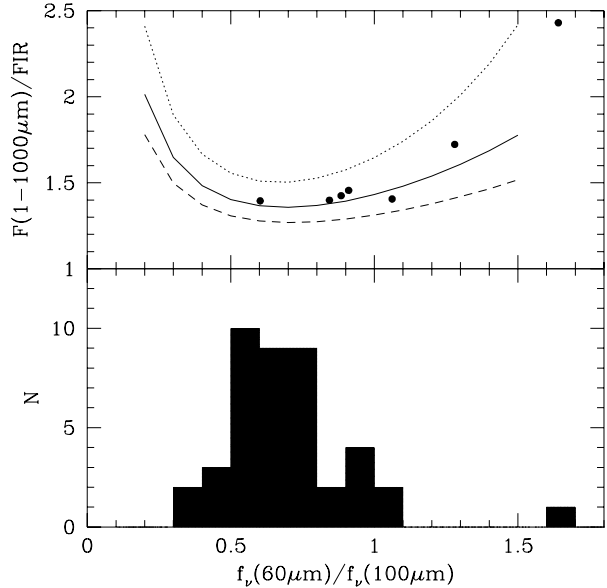,width=8.5cm}}}
\caption{The top panel shows the dust bolometric correction
BC(FIR)$_{\rm dust}$ for the non-AGN ultra luminous infrared galaxies in
the sample of Rigopoulou \etal\ (1996) plotted against FIR color.  The
dotted, solid, and dashed lines show the single temperature dust
emission models of Helou \etal\ (1988), for emissivity indices $n = 0$,
1, and 2 respectively.  The bottom panel shows the distribution of FIR
colors for local UV selected starbursts.
\label{f:bcfir}}
\end{figure*}

The bolometric dust correction ${\rm BC(FIR)_{Dust}}$ was estimated
empirically by directly calculating ${\rm BC(FIR)_{Dust}}$ for the
ultra-luminous infrared galaxies observed in the FIR-submm by
Rigopoulou, Lawrence, \&\ Rowan-Robinson (1996).  Two galaxies (Mrk~231
and IRAS05189-2524) were rejected from their sample because they have
strong AGN emission.  A smooth SED was drawn through the SED
observations of the remaining sources and integrated over $1\mu{\rm m} <
\lambda < 1.0$mm to determine $F_{\rm bol}$.  The individual ${\rm
BC(FIR)_{Dust}}$ measurements are shown in the top panel of
Fig.~\ref{f:bcfir}, as a function of FIR color $f_\nu(60\mu{\rm
m})/f_\nu(100\mu{\rm m})$.  For comparison the FIR colors of local UV
selected starbursts are shown in the bottom panel of Fig.~\ref{f:bcfir}.
We find that six of the seven remaining galaxies in the Rigopoulou
\etal\ sample have $\langle BC(FIR)_{Dust} \rangle = 1.47$ with a
dispersion of 0.13.  The seventh, 08572+3915 has $f_\nu(60\mu{\rm
m})/f_\nu(100\mu{\rm m}) = 1.64$, much warmer than the majority of the
UV selected starbursts, almost all of which have $0.3 < f_\nu(60\mu{\rm
m})/f_\nu(100\mu{\rm m})< 1.1$.

The top panel of Fig.~\ref{f:bcfir} shows that the empirical ${\rm
BC(FIR)_{Dust}}$ measurements are in good agreement with Helou's \etal\
(1988) dust emission models.  These models are very simple.  They treat
the dust distribution as emitting thermal radiation at a single
temperature, with emissivity $\propto \nu^n$.  We caution that more
realistic modeling of interstellar dust emission includes multiple
temperature emission with $n \approx 1$ to 2 (e.g. D\'{e}sert,
Boulanger, \&\ Puget 1990).  Figure \ref{f:bcfir} illustrates that
single temperature models with $n \approx 1$ provide a good
approximation to the dust emission of starbursts for the limited purpose
of calculating ${\rm BC(FIR)_{Dust}}$.  Adopting this model we have
calculated ${\rm BC(FIR)_{Dust}}$ for the local UV selected starbursts,
which tend to have somewhat cooler colors in the FIR than ultraluminous
infrared galaxies.  These have a median ${\rm BC(FIR)_{Dust}} = 1.37$,
with 80\%\ of the sample having $1.36 \leq {\rm BC(FIR)_{Dust}} \leq
1.44$.

We adopt ${\rm BC(1600)_\star} = 1.66 \pm 0.15$, and ${\rm
BC(FIR)_{Dust}} = 1.4 \pm 0.2$, yielding $B = 1.19 \pm 0.20$. Here the
``errors'' actually indicate the amount of leeway allowed by our model
calculations and limited measurements of starbursts, as described above.
Equation~\ref{e:irxlin} can now be written logarithmically as
\begin{equation}
\log(\irx) = \log(10^{0.4A_{1600}} -
1) + \log B,
\end{equation}
\begin{equation}
\log(\irx) = \log(10^{0.4A_{1600}}-1) + 0.076 \pm 0.044.\label{e:irxcal}
\end{equation}
Here we see the beauty of the approximation eq.~\ref{e:irx2}; it
allows the direct transformation of \irx\ to 1600\AA\ absorption as
shown on the right hand side of Fig.~\ref{f:uvir}.  

Our adopted model for the \irx\ - $\beta$ relationship is a simple
linear fit of $A_{1600}$ versus $\beta$.  To first order this is
what one would expect for absorption by a foreground screen of
dust.  The adopted fit, which is shown in Fig.~\ref{f:uvir}, is 
\begin{equation}
A_{1600} = 4.43 + 1.99\beta.\label{e:afit}
\end{equation}
It was determined from separate unweighted least squares fits of
$A_{1600}$ versus $\beta$, and $\beta$ versus $A_{1600}$, which were
then averaged.  The dispersion about this fit is 0.55 mag in $A_{1600}$,
0.28 in $\beta$ and 0.30 dex in $\log(\irx)$ (the latter including only
galaxies with $\beta > -2.23$).  Consequently the standard error in the
fit zeropoint is 0.08 mag in terms of $A_{1600}$ and 0.04 in terms of
$\beta$.  Since the scatter about the fit is significantly larger than
the errors, it is largely intrinsic.  Hence, a more sophisticated fit
(e.g.  a full $\chi^2$ fit with weighting by errors) will not improve
the quality of the fit.  Instead, improvement can only occur with an
increase in sample size.

This fit implies the spectral slope in the absence of dust absorption is
$\beta_0 = -2.23$.  This result is consistent with what is expected of
naked ionizing populations, which Leitherer \&\ Heckman (1995) show to
have $\beta_0 = -2.0$ to --2.6 (cf.\ Paper I).  In Table~\ref{t:alaws}
the slope of the absorption law, $dA_{1600}/d\beta$ derived here is
compared with other extinction and absorption laws.  The slope we derive
is in the middle of the range for the starburst absorption laws derived
in other studies.  In particular, it is between the values derived by
Calzetti \etal (1994) and Calzetti (1997).  Note that the Galactic
extinction law has a very large slope because the 2175\AA\ absorption
feature counteracts the reddening of $\beta$ to a large degree.

{\scriptsize
\begin{deluxetable}{llcl}
\tablecaption{1600\AA\ absorption law slopes\label{t:alaws}}
\tablehead{\colhead{law} & \colhead{type} &
\colhead{$dA_{1600}/d\beta$} & \colhead{reference} } 
\startdata
{\bf Starburst}  & {\bf absorption} & {\bf 1.99} & {\bf This work} \nl
Starburst  & absorption & 1.44 & Kinney et al.\ (1994) \nl
Starburst  & absorption & 1.97 & Calzetti et al.\ (1994) \nl
Starburst  & absorption & 2.30 & Calzetti (1997) \nl
Galactic   & extinction & 4.42 & Seaton (1979) \nl
LMC/30 Dor & extinction & 1.45 & Howarth (1983) \nl
SMC        & extinction & 1.05 & Bouchet et al.\ (1985) \nl
\enddata
\end{deluxetable} }

\subsection{Photometric versus spectroscopic $\beta$}

In Paper II we derived the following relationship between $\beta$ and
$(V_{606} - I_{814})_{\rm AB}$ color:
\begin{equation}
\beta_{\rm phot} = 3.23(V_{606} - I_{814})_{\rm AB} - 2,\label{e:bphot}
\end{equation}
where the color is in the ``AB'' system ($m_{\rm AB} = -48.6 -
2.5\log(f_\nu)$, where $f_\nu$ is the spectral flux density per
frequency interval in units of ${\rm erg\, cm^{-2}\, s^{-1}\,
Hz^{-1}}$).  This was derived from first principles using the central
wavelengths of the F606W and F814W WFPC2 filters and assuming a
featureless power law spectrum and negligible filter widths.  However,
the UV spectra of starbursts typically contain numerous absorption
features of both stellar and interstellar origin (Kinney \etal\ 1993).
Emission lines, even \Lya, are typically weak in starburst galaxies
(Kinney \etal, 1993).  While the spectroscopically defined $\beta$
avoids the lines, broad band filters cannot.  Since the concentration of
lines is particularly strong for $1216 < \lambda < 1700$\AA\ one may
expect an artificial reddening of the photometric $\beta$ at $z \approx
3$ when these features are redshifted into the middle of the F606W
passband.  In addition at $z \geq 2.8$, \Lya\ enters the F606W passband.
The Lyman forest strongly suppresses the flux bluewards of \Lya\ in such
high redshift objects, again reddening $\beta_{\rm phot}$.

\begin{figure*}
\centerline{\hbox{\psfig{figure=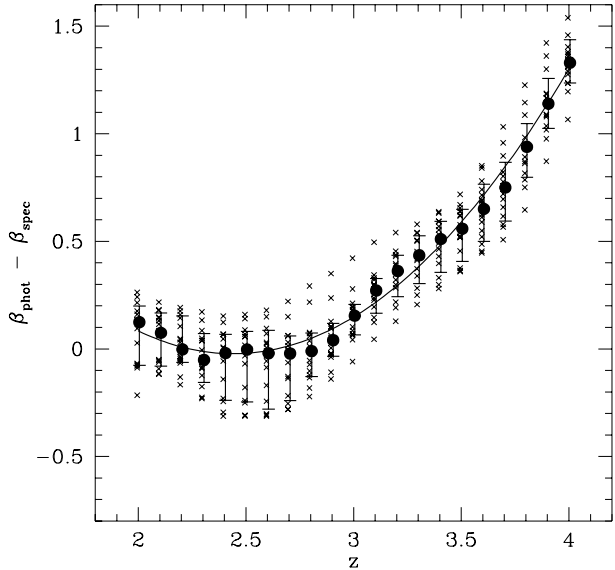,width=8.5cm}}}
\caption{The difference between photometric and spectroscopic $\beta$
plotted as a function of redshift.  The $\times$ marks indicate the
individual measurements of the 15 template spectra.  The circles
represent the median of the measurements and the errorbars the 16th
and 84th percentiles (i.e. $\approx 1\sigma$ errorbars).  The solid
line is the adopted polynomial calibration fit.
\label{f:betaphot}}
\end{figure*}

We have used IUE starburst spectra to quantify these effects and model
the calibration between photometric and spectroscopic $\beta$.  The
spectra we use were selected (by eye) to have high S/N and no
discernible break between LW and SW portion of spectrum.  This resulted
in a final sample of 15 galaxies, as indicated in Table~\ref{t:loc};
most of these were also used to calibrate the $A_{1600}$ versus $\beta$
fit.  Their spectroscopic $\beta$ were measured as described above.
These spectra were redshifted over the range $2 \leq z \leq 4$.  In
order to adequately account for the F606W filter width, the spectrum
blueward of \Lya\ was taken to be a pure power law (using the measured
$\beta$) normalized to the mean flux density at $\lambda = 1483 \pm 50$
\AA, and attenuated by the Lyman series forest which was modeled
following the algorithm of Madau (1995).  The $(V_{606} - I_{814})$
colors were then measured using SYNPHOT, yielding $\beta_{\rm phot}$.
Figure~\ref{f:betaphot} plots the difference between $\beta_{\rm phot}$
and the standard spectroscopic $\beta$ as a function of $z$.

Our calibration of the $\beta_{\rm phot} - \beta$
relationship is a quadratic polynomial fitted to the median points shown
in Fig.~\ref{f:betaphot}:
\begin{equation}
\beta_{\rm phot} - \beta = 3.22 - 2.66z + 0.545z^2.\label{e:dbz}
\end{equation}
The fit is a $\chi^2$ fit where the errors were taken to be the mean
difference between the median and the 16th and 84th percentiles for a given
$z$.  The fit has an rms dispersion of 0.06. The reduced $\chi^2$ is 0.19,
hence the model is adequate for the given amount of
scatter.  Combining this with eq.~\ref{e:bphot} yields:
\begin{equation}
\beta = 3.23(V_{606} - I_{814})_{\rm AB} - 5.22 + 2.66z
- 0.545z^2 \label{e:bphot2}
\end{equation}
which is now on the same scale as the $\beta - A_{1600}$ calibration
of eq.~\ref{e:afit}.

\subsection{Absolute star formation rate}

The transformation of UV luminosity to SFR is highly model dependent.
Hence, for most of this paper the emphasis will be on luminosity rather
than the SFR.  However, for comparison to other studies and to get a
feel for the physics behind the UV luminosity, it is often convenient to
transform it to a SFR. Our transformation is based on the population
models of Leitherer \&\ Heckman (1995).  We adopt a solar metallicity,
continuous star formation history population of duration 100 Myr, and a
Salpeter (1955) initial mass function (IMF) with mass limits of 0.1 to
100 \Msun. For these parameters a SFR of 1 \Msun\ yr$^{-1}$ produces a
1600\AA\ flux density of $7.86\, \times 10^{27}\, {\rm erg\, s^{-1}\,
Hz^{-1}}$ ($9.19\, \times 10^{39}\, {\rm erg\, s^{-1}\, \AA^{-1}}$), and
hence an absolute magnitude of --18.15 ABmag (--20.81 STmag); the
equivalent bolometric flux corresponds to $6.4\times 10^9\, L_\odot$.
For the same star formation history and IMF parameters, Bruzual \&\
Charlot (1993) stellar population models produce 8\%\ more flux at
1600\AA\ (see also comparison in Paper I). Not only is the choice of the
particular population model important in this calibration, so to are the
parameters of the IMF and the adopted star formation history.  For
example, if the lower mass limit of the IMF is raised to 1 \Msun\ then
the luminosity increases by a factor of 2.55 for a given SFR; if the
duration of star formation is decreased to 10 Myr (1 Myr), then the
luminosity decreases by a factor of 1.53 (12.9).

\section{Application to $U$-dropout galaxies in the HDF}\label{s:appl}

\subsection{Sample selection}

The above calibrations were applied to a sample of F300W ($U$) dropouts
selected from the HDF V2.0 catalog (Williams \etal\ 1996).  We will
compare the results from this sample to two other $U$-dropout samples
extracted from the HDF by Madau and collaborators - in M96 and Madau
\etal\ (1998; hereafter M98).  M96 lays out all selection criteria in
detail and lists all galaxies selected, and in addition it is the
pioneering work in this field.  M98 improves on M96 in several ways (see
below).  However, the results of M98 are model dependent, their
$U$-dropout selection criteria are not explicitly given, and their final
sample is not listed.  Hence we adopt M96 as our primary comparison, and
we model our sample selection on the criteria given therein.

\begin{figure*}
\centerline{\hbox{\psfig{figure=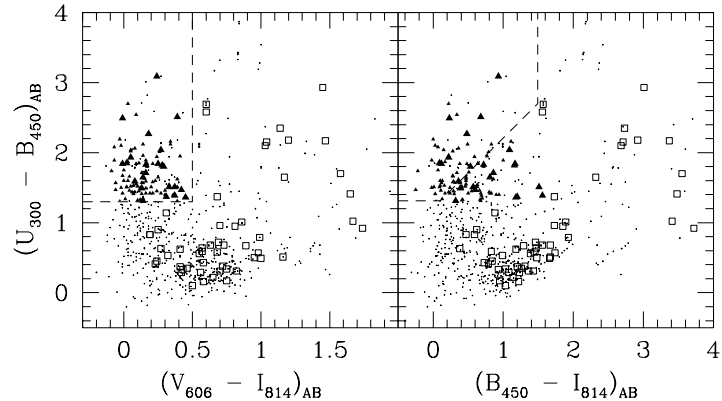,width=10cm}}}
\vspace{-4cm}
\caption{Two color diagrams of all objects (parents and daughters) in
the HDF V2.0 catalog (Williams \etal\ 1996) with $B_{450} \leq 26.79$
ABmag.  The dashed line in the left panel marks our adopted selection box
for $U$-dropouts, while that in the right panel marks the selection box
of M96.  The symbol correspondence is as follows: large triangles -
spectroscopically confirmed galaxies with $2 \leq z \leq 4$; squares -
galaxies with spectroscopic $z < 2$; small triangles - other galaxies
meeting our $U$-dropout selection criteria (i.e.\ they must be in the
box at left); dots - other objects not meeting our selection criteria.
Some of these latter objects are nevertheless in our selection box.
However they do not pass our simple parent-daughter separation
algorithm, and are not included (and hence are not highlighted) in order
to avoid double-counting.
\label{f:two2cd}}
\end{figure*}

Our sample selection differs from that of M96, primarily because we use
$(V_{606} - I_{814})_{\rm AB}$ rather than $(B_{450} - I_{814})_{\rm
AB}$ as a color selection criterion.  This is done because for
$U$-dropouts, the F450W filter is much more strongly affected by \Lya\
forest absorption than F606W.  In our model, selecting by $(V_{606} -
I_{814})_{\rm AB}$ places a fairly strong limit on $\beta$ and hence on
$A_{1600}$.  Figure~\ref{f:two2cd} shows the $(V_{606} - I_{814})_{\rm
AB}$, $(U_{300} - B_{450})_{\rm AB}$ two color diagram, and our
selection box.  Our selection criterion $(V_{606} - I_{814})_{\rm AB} <
0.5$ was chosen to encompass the strong $U$-dropout plume but avoid
relatively low-$z$ interlopers.  Our other selection criteria are
virtually identical to M96; $(U_{300} - B_{450})_{\rm AB} \geq 1.3$ is
the crucial $U$-dropout selection criterion, while $B_{450} \leq 26.79$
ABmag is the magnitude limit\footnote{Nominally, we have also adopted
$V_{606} \leq 28.0$ ABmag in our selection.  However, for our other
selection criteria no objects approach this limit.}.  Only sources in
the WF chips are selected.

Another important difference with M96 is that their color selection box
is not rectangular.  They have an additional constraint $(U_{300} -
B_{450})_{\rm AB} > (B_{450} - I_{814})_{\rm AB} + 1.2$, clipping the
bottom right corner out of the box.  This was introduced to avoid
regions that model calculation indicate may be populated by low redshift
galaxies.  However, this corner, while relatively sparsely populated in
the $(U_{300} - B_{450})_{\rm AB}$ versus $(B_{450} - I_{814})_{\rm AB}$
two color diagram, contains about \onethird\ the galaxies in the HDF
with measured $2 < z < 4$ (Dickinson 1998).  Since the clipped corner is
red in $(B_{450} - I_{814})_{\rm AB}$ the clipping biases against large
$A_{1600}$.  The color selection criteria of M98 are illustrated in
Madau (1997), where they were revised in order to incorporate more
galaxies with $2 < z < 4$.  Less of the bottom right corner is clipped
than in M96.  We have opted for a rectangular selection box in order to
minimize this bias and keep the color selection as simple as possible,
and yet be very close to the M96 selection box. A total of 100 sources
meet our selection criteria. These are listed in Table~\ref{t:udrop}.
Photometry for these galaxies can be found in Williams \etal\ (1996). In
comparison M96 selected 69 $U$-dropouts, and Madau (1997) note that
there are $\sim 100$ galaxies in their revised selection box. Dickinson
(1998) using more robust photometry, a somewhat different color
selection and a fainter limiting magnitude, isolates 187 $U$-dropouts in
the HDF.

{\scriptsize
\begin{deluxetable}{l l l l}
\tablecaption{List of $U$-dropouts selected in the HDF\label{t:udrop}}
\tablehead{} 
\startdata
2-35.0     & 2-689.2  & 3-389.0  & 4-317.0 \nl
2-76.0     & 2-726.0  & 3-504.0  & 4-363.0 \nl
2-80.0     & 2-741.0  & 3-510.0  & 4-382.0 \nl
2-82.0     & 2-751.0  & 3-515.11 & 4-389.0 \nl
2-122.0    & 2-793.0  & 3-550.0  & 4-445.0 \nl
2-127.0 & 2-802.11112 & 3-593.0  & 4-488.0 \nl
2-239.0    & 2-822.0  & 3-633.1  & 4-491.0 \nl
2-321.2    & 2-824.0  & 3-674.0  & 4-497.0 \nl
2-373.0    & 2-889.0  & 3-677.0  & 4-555.1 \nl
2-381.12   & 2-890.0  & 3-736.0  & 4-557.0 \nl
2-392.0    & 2-901.0  & 3-748.0  & 4-576.0 \nl
2-449.0    & 2-903.0  & 3-813.0  & 4-588.0 \nl
2-454.0    & 2-916.0  & 3-875.0  & 4-590.0 \nl
2-456.12   & 2-949.0  & 3-902.0  & 4-599.0 \nl
2-456.22   & 2-973.11 & 3-916.0  & 4-602.0 \nl
2-525.0    & 3-41.0   & 4-52.0   & 4-639.0 \nl
2-547.0    & 3-83.0   & 4-83.0   & 4-660.0 \nl
2-565.0    & 3-118.0  & 4-85.2   & 4-676.0 \nl
2-585.1    & 3-126.0  & 4-96.0   & 4-681.0 \nl
2-591.0    & 3-221.2  & 4-109.0  & 4-724.11 \nl
2-594.0    & 3-243.0  & 4-194.0  & 4-825.0 \nl
2-604.1    & 3-286.0  & 4-252.0  & 4-858.0 \nl
2-637.0    & 3-296.1  & 4-274.0  & 4-878.1 \nl
2-643.0    & 3-321.2  & 4-289.0  & 4-926.2 \nl
2-689.11   & 3-335.0  & 4-316.0  & 4-936.0 \nl
\enddata
\end{deluxetable} }

A small difference with M96 concerns how sub-structure is included. The
HDF catalog lists objects found with the FOCAS software (Jarvis \&\
Tyson 1981), which often splits a ``parent'' object into ``daughters''.
Both parents and daughters are listed in HDF V2.0.  M96 carefully
evaluate the image of each candidate $U$-dropout to determine whether it
is likely to apply to the whole parent object or the daughter(s).  Here,
in cases where both daughter(s) and corresponding parent meet our
selection criteria, only the parent object was included in our list of
$U$-dropouts. This could cause the integrated 1600\AA\ luminosity
density $\rho_{1600}$ to be overestimated.  This is because not all
daughters need satisfy the $U$-dropout selection criteria for the parent
to satisfy these same criteria.  To estimate the maximum magnitude of
this effect we considered the opposite extreme - when both parent and
daughter(s) meet the selection criteria only the (lowest level)
daughters that satisfy the selection criteria were included.  This
algorithm, results in $\rho_{1600}$ being lower than our adopted sample
by 16\%\ (24\% after absorption correction).  This algorithm will
underestimate $\rho_{1600}$ because some daughters, having the right
$z$ to be selected as $U$-dropouts, will fail the magnitude limit, or
because photometric errors (larger for the fainter daughters) may shift
some daughters out of the color selection box.  The ideal
parent-daughter separation will, therefore, lie somewhere between these
two extremes, and hence the effect on $\rho_{1600}$ is likely to be a
$\sim 10$\%\ effect.

\subsection{Application of calibration}

\begin{figure*}
\centerline{\hbox{\psfig{figure=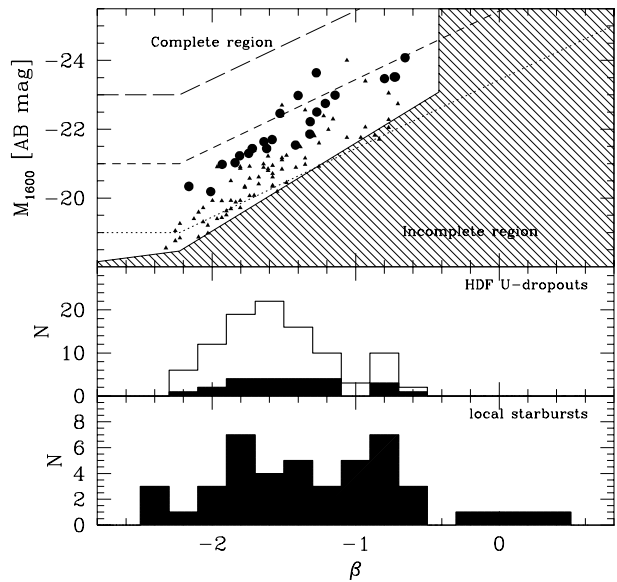,width=8.5cm}}}
\caption{\small The top panel shows the UV color $\beta$ verus
absorption corrected absolute magnitude $M_{1600,0}$ diagram for our
sample of $U$-dropouts.  Large circles mark spectroscopically
confirmed galaxies, while the rest of the $U$-dropouts are marked with
triangles.  The broken lines mark equal $M_{1600}$ without dust
absorption correction.  The hatched region marks the incomplete region
according to our selection criteria.  The slanted boundary is
calculated for galaxies at the limiting $B_{450} = 26.79$ ABmag with
$z = 2.75$. It follows from eqs.~\ref{e:afit} and \ref{e:bphot2} and
relates the $B_{450}$ limiting magnitude to the $V_{606}$ and
$I_{814}$ bands using $(B_{450} - I_{814})_{\rm AB} = 0.19 + 2.09 (V_{606} -
I_{814})_{\rm AB}$, which was found using a linear least squares fit to the
$U$-dropouts having $B_{450} \leq 26$ ABmag.  The middle panel shows
the $\beta$ distribution of these galaxies.  The solid histogram
includes only the spectroscopically confirmed $U$-dropouts.  For
comparison the $\beta$ distribution of the local starburst sample
(from Fig.~\ref{f:uvir}) is shown in the bottom panel.
\label{f:udrop}}
\end{figure*}

The top panel of Figure~\ref{f:udrop} shows the color ($\beta$)
absolute magnitude $M_{1600,0}$ of our HDF $U$-dropout sample.  Here,
$\beta$ was calculated from the $(V_{606} - I_{814})_{\rm AB}$ color listed in
the HDF V2.0 catalog (Williams \etal\ 1996) using eq.~\ref{e:bphot2};
where $z$ was taken from the compilation of Dickinson (1998) where
possible, otherwise we adopt an average value $\langle z\rangle =
2.75$.  This is the same $U$-dropout $\langle z\rangle$ adopted by
M96, and very close to $\langle z\rangle = 2.74$ of the galaxies in
our sample with measured $z$.  The apparent AB magnitude at $\lambda =
1600{\rm \AA}(1 + z)$ was interpolated between the $V_{606}$ and the
$I_{814}$ magnitudes.  This is effectively a $k$ correction. The $y$
axis shows the absolute AB magnitude at $\lambda_0 =1600$\AA\
corrected for UV absorption following eq.~\ref{e:afit}, while the
broken diagonal lines indicate constant $M_{1600}$ if there were no
$A_{1600}$ correction.  The hatched region shows the region that is
incomplete due to our selection criteria.  The lower selection limit
is slanted because of the $A_{1600}$ correction and because of color
terms in the $k$ correction and in the translation of the $B_{450}$
limiting magnitude into the $V_{606}$ and $I_{814}$ bands.

The middle panel of Fig.~\ref{f:udrop} shows the $\beta$ distribution of
the $U$-dropouts, while the bottom panel shows that of the calibrating
local sample for comparison.  The color distribution of the local sample
spans a larger range than the $U$-dropouts.  The two-sided
Kolmogorov-Smirnov test gives a 0.3\%\ chance that the two distributions
were randomly drawn from the same parent distribution.  These
differences are partially due to our $(V_{606} - I_{814})_{\rm AB} < 0.5$
selection limit which corresponds to $\beta \lesssim -0.42$
for $z = 2.75$, or equivalently $A_{1600} \lesssim 3.6$ mag.  The local
sample has no specific $\beta$ selection.  If we apply a $\beta <
-0.42$ limit to the local sample then the two-sided Kolmogorov-Smirnov
test gives a 3\%\ chance that the local and $U$-dropout sample were
randomly drawn from the same parent distribution; i.e.\ the two samples
are more similar when similar $\beta$ selection criteria are applied.
The $U$-dropout sample is {\em bluer\/} than the local sample having
$\beta < -0.42$.  The 20th percentile, median and 80th percentile of the
$U$-dropout distribution are --1.9, --1.6, and --1.2, while for the
($\beta$ limited) local sample they are --1.8, --1.4, and --0.8,
respectively.  One shouldn't read too much into this difference, in that
the two samples were selected quite differently.  The $U$-dropouts have
well defined magnitude and color limits in the rest-frame UV while the
local sample is composed of galaxies observed with IUE for a wide range
of programs with heterogeneous initial selection criteria.

\begin{figure*}
\centerline{\hbox{\psfig{figure=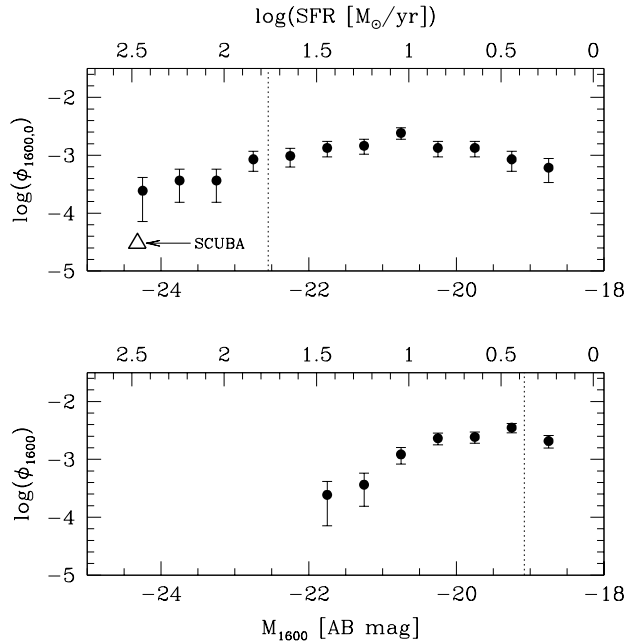,width=8.5cm}}}
\caption{Luminosity function ($\phi$) at $\lambda = 1600$\AA\ of our
sample of $U$-dropouts, with (top) and without dust absorption
corrections.  The error bars show the $\sqrt{N}$ uncertainties. The
dotted line shows the completeness limit.  Data to the left of these
lines are complete, to the right they are incomplete.  The top axis of
each plot converts $M_{1600}$ to star formation rate.  The units for
$\phi$ are ${\rm mag^{-1}\, Mpc^{-3}}$.  The space density and
luminosity of SCUBA sources, as estimated by Lilly \etal\ (1998), is
indicated in the upper panel with a triangle.
\label{f:lf}}
\end{figure*}

Luminosity functions at 1600 \AA, $\phi_{1600}$ are shown for our $U$
dropout sample in Fig.~\ref{f:lf}.  Two cases are shown: with
($\phi_{1600,0}$) and without ($\phi_{1600}$) the $A_{1600}$ correction.
The shape of $\phi_{1600}$ agrees well with that given by Dickinson
(1998) for $M_{1600} \lesssim -19$ ABmag.  The drop-off in the last bin
of our $\phi_{1600}$ is due to incompleteness.  Dickinson's luminosity
function extends further because of his superior photometric technique
and consequent fainter limiting magnitude.  Our luminosity function is
0.3 dex higher than that displayed by Dickinson because he renormalizes
to match the luminosity function of $U$-dropouts selected from a larger
ground based survey (Steidel \etal\ 1998).  This factor of two
difference probably reflects the field to field variation in
$\rho_{1600}$.

The 1600\AA\ luminosity functions extend well beyond the range given for
a sample of local UV selected galaxies by Buat \&\ Burgarella (1998).
Their luminosity function shows a distinct break at $L_{\rm bol} \sim
10^{10}\, L_\odot$, corresponding to $M_{1600,0} = -18.7$ ABmag; a SFR
of 4 \Msun\ year$^{-1}$.  The brightest galaxies in our $U$-dropout
sample have $M_{1600} = -21.7$ ABmag, $M_{1600,0} = -24.1$,
corresponding to a SFR of 19, 220 \Msun year$^{-1}$ with and without
absorption correction, respectively.  This is further confirmation of
the assertion that $U$-dropouts are much more luminous than local UV
selected starbursts.  However, there are local starbursts with similarly
high SFRs.  These are the ultra-luminous infrared galaxies (Sanders \&\
Mirabel, 1997) which are selected in the far infrared.

From direct summation we derive the total absorption corrected UV
luminosity of our selected $U$-dropouts to be $\Sigma L_{1600,0} = 3.8\,
\times 10^{34}\, {\rm erg\, s^{-1}}$.  Assuming that the
redshift range of $z = 2.0$ to 3.5 is uniformly sampled, then the volume
containing the $U$-dropouts is $1.64 \times 10^4\, {\rm Mpc^3}$ (M96),
and the corresponding volume emissivity is $\rho_{1600,0} = 1.4 \pm 0.3
\times 10^{27}\, {\rm erg\, s^{-1}\, Hz^{-1}\, Mpc^{-3}}$ ($\rho_{\rm
SFR} = 0.18\, \Msun\, {\rm yr^{-1}\, Mpc^{-3}}$).  The error quoted here
is the quadratic sum of the zeropoint errors from equations
\ref{e:irxcal}, \ref{e:afit} and \ref{e:bphot2}, and the sampling error
determined from the bootstrap technique (e.g.\ Babu \&\ Feigelson
1996).  This $\rho_{1600,0}$ is actually a lower limit for the volume
sampled by the HDF because we have made no completeness correction.
This raises a difficult issue.  The standard practice is to estimate the
completeness from the luminosity function, without dust absorption.
Doing so, Madau (1997) estimates a completeness correction factor of
1.45.  Figures~\ref{f:udrop} and \ref{f:lf} show that in the presence of
dust, the limiting $M_{1600,0}$ is not uniform.  Rather it is much
shallower for highly reddened galaxies than for lightly reddened
sources.  Hence the completeness correction could be higher.  In
addition there may be field to field differences in $\rho_{1600,0}$, due
to large scale clustering.  Such variations could be as high as a factor
of 2 (Dickinson, 1998).

How accurate is our $U$-dropouts selection criteria for selecting
high-$z$ galaxies?  Our sample includes all but three of the
spectroscopically confirmed HDF galaxies with $z > 2$ as compiled by
Dickinson (1998).  The galaxy 1-54.0 was missed because it appeared 
on the PC1 chip rather than one of the WF chips; 3-512.0 has $z = 4.02$
and is $B$-dropout and hence too red to be a $U$-dropout; and 3-577.0
was not included because it is too faint for inclusion in our catalog.
The remaining 23 spectroscopically confirmed high-$z$ galaxies account
for 47\%\ of $\rho_{1600,0}$.  We cross-checked our $U$-dropout selection
against a list of all redshift measurements in the HDF (kindly made
available by Mark Dickinson).   None of our selection  corresponds to
galaxies with known $z < 2$.  Dickinson (1998) notes a 90\%\ success
rate in finding high-$z$ galaxies in the appropriate redshift range
using the $U$-dropout technique on their ground based sample.  While the
photometric bands and selection criteria are somewhat different from
ours, this suggests that the contribution of low-$z$ interlopers to
$\rho_{1600,0}$ should not be much higher than 10\%. 

\subsection{The net effect of dust absorption}

In this paper we are primarily interested in the effects of dust on the
high-$z$ UV emissivity.  If we calculate $\Sigma L_{1600}$ and
$\rho_{1600}$ without correcting for dust absorption, the quantities are
a factor of $5.4 \pm 0.9$ lower than the above estimates,
e.g. $\rho_{1600} = 2.7 \pm 0.2 \times 10^{26}\, {\rm erg\, s^{-1}\,
Hz^{-1}\, Mpc^{-3}}$.  This is the net effect of dust absorption on {\em
our sample\/} of $U$-dropouts.  For comparison, M96 give $\rho_{1600} =
1.6 \times 10^{26}\, {\rm erg\, s^{-1}\, Hz^{-1}\, Mpc^{-3}}$ for their
sample of $U$-dropouts.  The samples are identical in their limiting
magnitudes, so the difference in the uncorrected emissivities is due to
the sample selection.  In particular, Fig.~\ref{f:two2cd} shows that the
portion of $\Sigma L_{1600}$ missing from the M96 sample is contained
within the corner clipped from their otherwise rectangular color
selection box.  This corner contains preferentially reddened galaxies,
hence this clipping biases against large $A_{1600}$.  Relative to
$\rho_{1600}$ estimated by M96 our emissivity is a factor of 9.0 larger.
We reiterate that the actual absorption-corrected $\rho_{1600,0}$ will
be even higher than our estimates because we have not corrected for
completeness, and we are not sensitive to galaxies having $A_{1600} >
3.6$ mag.

More recently, Madau (1997) and M98 now estimate an uncorrected
$\rho_{1600} = 2.6 \times 10^{26}\, {\rm erg\, s^{-1}\, Hz^{-1}\,
Mpc^{-3}}$ for the volume sampled by the HDF $U$-dropouts, only 4\%\
less than our estimate without dust correction.  It is higher than in
M96 because, as noted above, they adopt significant completeness
corrections (a factor of 1.45) and use revised $U$-dropout selection
criteria that clip less of the bottom-right corner in the $(U_{300} -
B_{450})_{\rm AB}$ vs.\ $(B_{450} - I_{814})_{\rm AB}$ plane.

The 1600\AA\ dust absorption factor, $5.4 \pm 0.9$ (equivalent to
$A_{1600} = 1.8 \pm 0.2$ mag), derived here for the $U$-dropouts is
lower than our original estimate of a factor of 15 (Paper II), albeit
within the factor of three uncertainty we originally estimated.  Sawicki
\&\ Yee (1997) using the Calzetti (1997) absorption law, derived
similarly high dust absorption factors of 15 to 20.  Our new estimates
are somewhat lower for a combination of reasons.  Firstly, we now fit
the \irx\ - $\beta$ relation instead of adopting an absorption law
(Calzetti \etal\ 1994) that is unconstrained by FIR data.  Secondly, in
Paper II we adopted the median absorption of the spectroscopically
confirmed $U$-dropouts.  Because of the $M_{1600,0} - \beta$ correlation
(Fig.~\ref{f:udrop}) this biases the net absorption.  Finally, when
deriving $\beta$ from $(V_{606} - I_{814})_{\rm AB}$ we now add a $z$
dependent correction (eq.~\ref{e:bphot2}) to remove Lyman-forest and
Lyman edge reddening, thus resulting in bluer colors and a smaller
derived $A_{1600}$.  

The uncertainty in our new estimate of the mean dust absorption factor
of $U$-dropouts amounts to 16\%, and includes zeropoint and sampling
errors.  It is likely to underestimate the true uncertainty since it
does not include field to field variations.  M98 and
Pettini \etal\ (1998) estimate 1500\AA\ dust absorption factors of about
three.  Thus the difference between these groups and ours has narrowed
considerabley, and is probably within realistic errors on the dust
absorption factor.  However we reiterate our caution: $5.4 \pm 0.9$
represents a lower limit to the true mean 1600\AA\ dust absorption
factor of galaxies at $z \approx 3$.  As discussed in \S\ref{ss:scuba},
true 1600\AA\ dust absorption factors on the order of 10 are still not
ruled out.

Buat \&\ Burgarella (1998) using a method very similar to ours, derive a
somewhat lower mean absorption of 1.2 mag at $\lambda = 2000$\AA\ for a
{\em local\/} sample of starbursts selected to be visible in both the UV
and FIR.  The difference probably is largely due to the lower luminosity
of the galaxies in their sample, which has a median $L_{\rm bol} \approx
10^9\, L_\odot$ (SFR $\approx 0.16\, \Msun\, {\rm year}^{-1}$).  The
$U$-dropout sample has a median $L_{\rm bol} = 8.5\times 10^{10}\,
L_\odot$ (SFR $\approx 13 \Msun\, {\rm year^{-1}}$), nearly a factor of
100 higher.  Because of the $M_{1600,0} - \beta$ correlation
(Fig.~\ref{f:udrop}), we expect the local sample to have a lower average
amount of dust absorption (cf.\ Heckman \etal\ 1998).

In summary, our value for the dust absorption corrected UV emissivity
$\rho_{1600,0} = 1.4 \times 10^{27}\, {\rm erg\, s^{-1}\, Hz^{-1}\,
Mpc^{-3}}$ at $\langle z\rangle = 2.75$ is a factor of 9.0 higher than
the uncorrected value given by M96 and a factor of 5.4 times higher than
the uncorrected value given by M98.  The increased UV
emissivity relative to M96 is due to a combination of an improved
$U$-dropout selection (factor of 1.7), and correction for dust
absorption (factor of 5.4).

\section{Tests of the starburst hypothesis}\label{s:test}

Our determination of the UV opacity in high-$z$ galaxies relies on the
similarity of the $U$-dropouts to local starburst galaxies; the
intrinsic SEDs of the two groups must be similar for our method to work.
In \S\ref{s:comp} we reviewed how the overall spectral properties of the
two samples are similar.  Here we concentrate on recent observational
studies of high-$z$ galaxies that are especially relevant to estimating
UV opacities.  

\subsection{Rest frame optical emission lines}

If the high-$z$ galaxies are intrinsically redder in the UV than local
starbursts (i.e.\ $\beta_0 > -2.2$), then we would overestimate
$A_{1600}$ and hence $\rho_{1600,0}$.  This could occur if the high-$z$
galaxies are dominated by {\em older\/} stellar populations (which would
have a larger $z$ of formation) or, if for some reason, their stellar
IMF were deficient in the highest mass stars.  Rest-frame optical
emission lines present an ideal test for this hypothesis.  Starbursts
are by nature dominated by highly ionizing populations.  Indeed the
calibrating IUE sample used in \S\ref{s:cal} were classified as
starbursts primarily by the presence of intense narrow optical emission
lines, arising in the ionized medium surrounding the starbursts.  Hence,
the ratio of line flux to flux density $f_{1600}$ is a measure of the
color from below the Lyman-limit to 1600\AA, modulated by dust
absorption.

\begin{figure*}
\centerline{\hbox{\psfig{figure=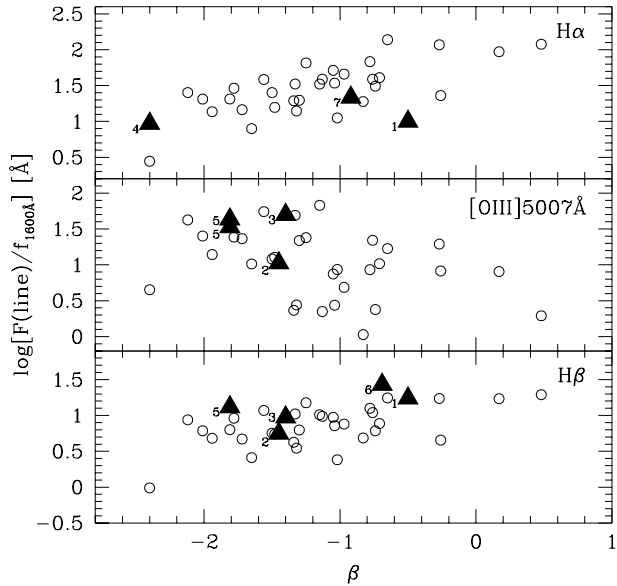,width=8.5cm}}}
\caption{Line flux to UV flux density ratio compared to UV spectral
slope.  The open circles show a local sample of UV selected
starbursts, using data originally published by Storchi-Bergman \etal\
(1995) and McQuade \etal\ (1995).  The labelled triangles represent
seven $U$-dropouts observed by Pettini \etal\ (1998), Wright \&\
Pettini (1998); Bechtold \etal\ (1997) and Bechtold \etal (1998).  The
label - galaxy name correspondence is (1) MS1512-cB58; (2) Q0000-D6;
(3) Q0201-C6; (4) Q0201-B13; (5) B2 0902-C6; (6) DSF 2237a-C2; and (7)
A2218-A384.  These data have been corrected for the redshift
stretching of the spectral energy distribution.
\label{f:uvem}}
\end{figure*}

For $z = 2 - 3.5$ rest-frame optical emission lines are redshifted into
the observed-frame near infrared.  With careful, and very long exposures,
the brightest optical lines are detectable in the brightest $U$-dropout
galaxies with four meter class telescopes.  Detections and fluxes have
been reported by several groups including Pettini \etal\ (1998), Wright
\&\ Pettini (1998), Bechtold \etal\ (1997), and Bechtold \etal\
(1998). The detections correspond to $U$-dropouts discovered from the
ground or in the HDF, as well as other sources that meet the photometric
criteria for being $U$-dropouts.  Figure~\ref{f:uvem} shows the line
fluxes of these sources normalized by the rest frame 1600 \AA\ flux
density $f_{1600}$ of the sources compared to $\beta$.  Both $f_{1600}$
and $\beta$ were estimated from published photometry, where possible
(otherwise from published spectra).  The pseudo-equivalent widths
$F({\rm line})/f_{1600}$ have been corrected for Galactic extinction and
cosmological stretching only.  No attempt has been made to correct for
internal absorption.  Results for three different lines are shown -
\Hbeta, \Halpha, and \forbid{O}{iii}{5007\AA}.  In cases where the
\forbid{O}{iii}{4959\AA} line was detected, it was multiplied by 3.0 to
convert it to an \forbid{O}{iii}{5007\AA} flux. A sample of local
starburst galaxies observed by Storchi-Bergmann \etal\ (1995) and
McQuade, Calzetti, \&\ Kinney (1995) is shown for comparison.  The
relevant data is listed in Table~\ref{t:loc}.  This UV selected sample
has a high degree of overlap with the sample used in \S\ref{s:cal}.

Figure~\ref{f:uvem} shows that the detected $U$-dropouts have line flux
to UV flux density ratios consistent with local starburst galaxies.  The
one exception is the \Halpha\ observation of the lensed galaxy
MS1512-cB58 by Bechtold \etal\ (1997).  Wright \&\ Pettini (1998) have
also detected this galaxy in \Hbeta\ emission, and the corresponding
\Hbeta\ pseudo-equivalent width is well within the scatter of points for
local starbursts as shown in the bottom panel of Fig~\ref{f:uvem}.  The
discrepancy between the \Halpha\ and \Hbeta\ measurements may be due to
the lensed nature of the source, or perhaps may reflect the difficulty
of the \Halpha\ observations which were obtained at $\lambda = 2.443\,
\mu{\rm m}$, the edge of the $K$ band window.  At this stage, none of
the groups report non-detections in emission lines of known
$U$-dropouts.  We conclude that $U$-dropouts are ionizing sources
to the same degree as local starbursts, and hence do not have an
intrinsically redder continuum than the local sample.  

\subsection{Far infrared and radio continuum observations\label{ss:rc}}

If there is significant UV extinction of galaxies at $z \approx 3$, then
perhaps the dust can be detected directly in the rest frame FIR or
indirectly in the radio via the radio-FIR correlation (Helou 1991). Two
groups have recently claimed (marginal) detections of sources in and
near the HDF in these wavelength regimes some of which which they
attribute to $z \approx 3$ star forming galaxies.  Here we show that the
number and fluxes of these marginal detections are consistent with the
expectations of our dust absorption model.

The sub-mm regime, corresponding to the rest frame FIR, is a promising
window to look for high-$z$ galaxies since the ``negative
$k$-corrections'' for dust emission can cancel out the effects of
fade-out with distance, making dusty galaxies easier to detect with
increasing $z$ (Guiderdoni \etal\ 1997).  Several groups have now
obtained deep sub-mm imaging with SCUBA\footnote{Sub-mm Common User
Bolometer Array} on the 15m JCMT\footnote{James Clerk Maxwell Telescope}
(e.g.\ Smail, Ivison \&\ Blain, 1997; Barger \etal\ 1998; Eales \etal\
1998; Hughes \etal\ 1998; Lilly \etal\ 1998).  All report 850\micron\
detections at the mJy level that they largely attribute to galaxies
having $z > 1$.  Indeed one of the SCUBA detections, SMM02399-0136, has
a clear optical counter-part and spectroscopic determination of $z=2.8$
(Ivison \etal\ 1998).  However, Lilly \etal\ (1998) note that a sizable
fraction of these sources may have $z < 1$.  Furthermore, the emission
line spectrum of SMM02399-0136 is indicative of a Seyfert 2 galaxy,
cautioning us that not all of the dust seen by SCUBA need be warmed by
star-formation alone.

The study of Hughes \etal\ (1998) is particularly relevant because it is
of a field centered on the HDF.  They report five sources above their
detection limit of 2mJy. Since this limit is set by source confusion
(beam FWHM$\approx 15''$), none of their optical identifications is
totally secure.  They attempt to improve the accuracy of their
identifications using photometry, including photometric redshifts.  They
claim that 4/5 of their detections are best identified with sources
having $z \gtrsim 2$.  The brightest of these has $f_{\rm 850\mu{m}} = 7
\pm 2$ mJy.

In Table~\ref{t:rad} we present $f_{\rm 850\mu{m}}$ estimates for HDF
$U$-dropouts based on our dust absorption model, and the optical
photometry of Williams \etal\ (1996).  To do this we used
eqns.~\ref{e:irxcal} and \ref{e:afit} to estimate the rest frame
$F_{\rm FIR}$ from the rest frame $F_{1600}$ and $\beta$ estimates.
$F_{\rm FIR}$ was then converted to $f_{\rm 850\mu{m}}$ using the FIR
SED of the prototypical IR luminous starburst Arp~220 as modeled by a
spline fit through the the measurements of Rigopoulou \etal\ (1996)
and Klaas \etal\ (1997).  Only the galaxies with the brightest
predicted $f_{\rm 850\mu{m}}$ values are listed.  The table is
arranged in order of decreasing predicted $f_{\rm 850\mu{m}}$.  The
corresponding rest frame 1600\AA\ flux densities (uncorrected for dust
absorption) do not come out to be similarly sorted.  This is because
it is the dust reprocessed flux that is important. Since our adopted
$U$-dropout sample is color limited (i.e.\ $A_{1600} \lesssim 3.6$
mag), the brightest predicted infrared fluxes correspond to the
brightest and reddest $U$-dropouts.  The UV absorption $A_{1600}$
listed in this table is calculated from the $(V_{606} - I_{814})_{\rm AB}$ as
outlined in \S\ref{ss:bacal}.

{\scriptsize
\begin{deluxetable}{l r c c c c c c c}
\tablecaption{HDF $U$-dropout FIR and radio flux predictions\label{t:rad}}
\tablehead{ & & & &\colhead{$\lambda$:} & 
\colhead{$(z+1)1600$\AA} & \colhead{$850\mu$m} & \colhead{3.5cm} & \colhead{3.5cm} \nl
& & & & & \colhead{Observed} & \colhead{\small Predicted} & \colhead{\small Predicted} & \colhead{\small Observed} \nl
\colhead{Name} & \colhead{$z$} &
\colhead{$I_{814}$} & \colhead{$\beta$} & \colhead{$A_{1600}$} &
\colhead{$f_\nu$} & \colhead{$f_\nu$} & \colhead{$f_\nu$}
& \colhead{$f_\nu$} \nl
& & \colhead{\small [ABmag]} & & \colhead{\small [mag]} &
\colhead{\small [$\mu$Jy]} & \colhead{\small [mJy]} & \colhead{\small [$\mu$Jy]} & \colhead{\small [$\mu$Jy]} }
\startdata
~4-52.0\tablenotemark{a}& 2.931 & 24.06 & --0.66 & 3.05 & 0.63 &   1.8 & 3.7 & $6.3 \pm 2.6$\nl
4-878.1            & (2.75)& 23.28 & --1.06 & 2.32 & 1.35 &   1.6 & 3.7 & \nodata \nl
4-639.0            & 2.591 & 24.16 & --0.73 & 2.92 & 0.52 &   1.1 & 2.8 & \nodata \nl
2-585.1            & 2.002 & 23.50 & --0.73 & 2.93 & 0.76 &   1.1 & 4.7 & \nodata \nl
4-555.1            & 2.803 & 23.33 & --1.27 & 1.93 & 1.39 &   1.1 & 2.3 & \nodata \nl
4-445.0            & 2.268 & 23.75 & --0.80 & 2.79 & 0.69 &   1.0 & 3.5 & \nodata \nl
4-382.0            & (2.75)& 24.10 & --0.95 & 2.52 & 0.61 &   0.9 & 2.2 & \nodata \nl
2-127.0            & (2.75)& 24.72 & --0.79 & 2.81 & 0.33 &   0.7 & 1.6 & \nodata \nl
2-449.0\tablenotemark{b}& 2.008 & 23.42 & --1.14 & 2.17 & 1.01 &   0.6 & 2.7 & $7.0 \pm 2.0$\nl
4-858.0            & 3.216 & 24.06 & --1.40 & 1.70 & 0.78 &   0.6 & 1.0 & \nodata \nl
\enddata
\tablenotetext{a}{Source 3647+1255 in Richards et al.\ (1998)}
\tablenotetext{b}{Source 3648+1416 in Richards et al.\ (1998)}
\end{deluxetable} }

None of these galaxies have predicted $f_{\rm 850\mu{m}} > 2$ mJy,
consistent with the fact that none were detected by Hughes
\etal. Likewise, none of the claimed optical counterparts of $f_{\rm
850\mu{m}}$ detections meet our criteria of being $U$-dropouts.  One is
clearly a $z \approx 1$ galaxy.  Three are fainter than our adopted
inclusion limit.  The remaining galaxy HDF850.3 is in the PC1 chip and
identified as 1-34.2.  Its $(V_{606} - I_{814})_{\rm AB} = 0.33$ is
within our selection limits, but its $(U_{300} - B_{450})_{\rm AB} =
1.08$ color is marginally too blue to be considered a $U$-dropout, but
consistent with a photometric $z \approx 1.95$ estimated by Mobasher
\etal\ (1996).

The non-intersection of our $U$-dropout catalogue and the 850$\micron$
sample of Hughes \etal\ suggests that the sub-mm detections correspond
to infrared luminous starbursts or AGN that are so dust enshrouded that
they are not selected as $U$-dropouts either because they are too
extincted and/or because they are too reddened.  The Hughes \etal\
identification of two 850\micron\ sources (HDF850.1, HDF850.2) with
fairly blue galaxies ($(V_{606} - I_{814})_{\rm AB} < 0.5$) also
suggests that the $\beta$-\irx\ calibration breaks down for heavily
extincted infrared-luminous starbursts.  Further work is needed to
confirm these identifications and secure reliable redshifts.

The remarkably tight radio - FIR correlation (Helou 1991) provides an
indirect means to probe the FIR emission, and hence dust emission of
starbursts.  It is best defined at $\nu = 1.4$ GHz (21cm; e.g.\
Sanders \&\ Mirabel, 1997).  The radio emission of a wide range of
star forming galaxies, including starbursts, is dominated by
synchrotron emission at this frequency and up to $\sim 30$ GHz.  Hence
the correlation can be used to infer FIR fluxes from radio
observations down to rest wavelengths of about 1 cm. Do we expect this
correlation to hold at high-$z$? Inverse Compton scattering from
cosmic microwave background photons provides an alternative mechanism
to synchrotron emission for cooling the relativistic electrons.  The
energy density of these photons will scale as $(1 + z)^4$.  However it
is the total energy density of photons that is important for inverse
Compton cooling.  Even at $z \approx 3$ the energy density of cosmic
microwave background photons will be an order of magnitude lower than
internally produced energy density of starbursts having typical
bolometric surface brightnesses of $S_e = 3 \times 10^{10}\, {\rm
L_\odot\, kpc^{-2}}$ (e.g.\ Paper II) at $z = 3$.  Since synchrotron
emission clearly occurs in the face of intense internally produced
emission in low-$z$ galaxies, it does not seem likely that inverse
Compton cooling will quench it for high-$z$ galaxies where the total
energy density is only marginally higher.

Richards \etal\ (1998) present a deep imaging survey of the HDF and
surrounding areas obtained with the VLA at 3.5 cm ($\nu = 8.46$ GHz).
Their $5\sigma$ detection limit is $9\, \mu$Jy, and their $3.5\sigma$
marginal detection limit corresponds to $6.3\, \mu$Jy. They marginally
detect two HDF $U$-dropouts, 4-52.0 and 2-449.0 as listed in the HDF
V2.0 catalog.  While the detections are provisional, Richards \etal\ are
confident that they are not spurious because they also are detected in a
1.4 GHz image of the HDF.

The logarithmic ratio of $f_\nu$ flux densities in the rest frame radio
(1cm $= 30$ GHz) and UV (1600\AA\ $= 1.9$ PHz) and  can be written as:
\begin{equation}
\log\left( \frac{f_{\rm 1cm}}{f_{\rm 1600\AA}} \right) =
2.78 - q_\nu({\rm 1cm}) - \log(10^{0.4 A_{1600}} - 1),
\end{equation}
where $q_\nu$ is the radio-FIR logarithmic ratio:
\begin{equation}
q_\nu = \log\left\{\frac{F_{\rm FIR}[{\rm 10^{-23}\, erg\, cm^{-2}\,
s^{-1}}]}{3.75 \times 10^{12}\,[{\rm Hz}]}\right/ f_\nu [{\rm Jy}] \left\}
\frac{}{} \right..
\end{equation}
At $\lambda =21$ cm, $q_\nu({\rm 21cm}) = 2.35$ (Sanders \&\ Mirabel,
1997).  Since starbursts typically have radio spectral slopes $\alpha
\approx -0.7$ (where $f_\nu \propto \nu^\alpha$), we expect $q_\nu({\rm
1cm}) = 3.26$.  For the prototypical starburst galaxy M82, $q_\nu({\rm
1cm}) = 3.16$ is derived from the data listed by Carlstrom \&\ Kronberg
(1991).  This is somewhat lower than expectations because free-free
emission is starting to make a contribution to the 1cm flux of M82
(Carlstrom \&\ Kronberg 1991).  We adopt a rest frame $q_\nu({\rm 1cm})
= 3.2$ and calculate the expected $f_\nu$ at 3.5cm ($f_{\rm 3.5 cm}$) of
the $U$-dropouts in our HDF sample using the above formulation, and
modest (factor of a few at most) ``$k$'' corrections to transfer from
the rest frame wavelengths of 1600\AA\ and 1cm, and the observed frame
$I_{814}$ and $\lambda = 3.5$ cm observations.  Table~\ref{t:rad} lists
the results of these calculations for the expected 10 sub-mm brightest
$U$-dropouts.  Note that the $f_\nu$ ranking of the radio predictions is
different from that of the FIR predictions.  This is because the $k$
corrections in the two wavelength regimes work in opposite direction
with redshift at $z \approx 3$.

The sources in Table~\ref{t:rad} include the two detections of Richards
\etal\ (1998).  Note that all of the galaxies in this table have
predicted $f_{\rm 3.5cm}$ below the formal detection limits of Richards
\etal.  However the top five $f_{\rm 3.5cm}$ estimates are not far below
their marginal detection limit, and include these two sources.  In such
cases we expect the fortuitous coincidences of uncertainties to allow
some positive detections.  These uncertainties include the measurement
bias mentioned above, the flux uncertainties listed by Richards \etal\
($\sim$0.1 dex uncertainty), the intrinsic scatter about the radio-FIR
relationship ($\sim 0.3$ dex uncertainty; Devereaux \&\ Eales 1989), and
the random scatter about the $A_{1600}$ versus $\beta$ fit ($\sim$0.1
dex uncertainty).  The total random uncertainty is $\sim 0.3$ dex or a
factor of $\sim 2$.  Hence random errors alone can account for the
difference in the predicted and observed $f_{\rm 3.5cm}$ in the two
galaxies detected by Richards \etal.  Why aren't these galaxies also
detected in the rest frame FIR?  Scaling our predictions by the $f_{\rm
3.5cm}$ observed by Richards \etal, we then expect $f_{\rm 850\mu{m}} =
3.2 \pm 1.3, 1.6 \pm 0.5$ mJy for 4-52.0 and 2-449.0 respectively.  The
latter is below the Hughes \etal\ detection limit, while the former is
above it but by less than the flux uncertainty, and so could easily be
missed.  We conclude that the Richards \etal\ radio detections are
consistent with the Hughes \etal\ SCUBA non-detections.

Our calculations show that $U$-dropouts in the HDF are just at the
detection limit of state of the art deep rest frame FIR and radio
observations.  Somewhat deeper observations in these bands of either the
HDF or other fields containing $U$-dropouts (preferably the brightest
and reddest ones) have the potential for directly testing the $\beta$
-- \irx\ relationship and verifying the existence of the radio-FIR
correlation at $z \approx 3$.  Hence they could be used to directly test
the form of the UV absorption law.  For example, the SMC law absorption
model adopted by Pettini \etal\ (1997) would predict $f_{\rm 850\mu{m}}
< 0.4$ mJy, and $f_{\rm 3.5cm} < 1\mu$Jy for all the $U$-dropouts in our
sample.  Hence mJy level imaging at 850\micron\ and $\mu$Jy level radio
continuum imaging should detect the brightest $U$-dropouts if our
absorption model is correct, but would not detect any if lightly
absorbed models were correct.

\section{Discussion}\label{s:disc}

\subsection{Color-luminosity correlation}

The color-magnitude diagram of HDF $U$-dropouts, Fig.~\ref{f:udrop},
shows that the highest luminosity galaxies, with or without $A_{1600}$
correction, tend to be red.  Similarly the bluest galaxies tend to have
the lowest luminosity.  That is, there is a color-luminosity correlation
(also noted by Dickinson, 1998, private communications).  The $\beta$ -
luminosity correlation is similar to that seen in local starburst
galaxies (Heckman \etal\ 1998).  As is the case for local starbursts,
this correlation is not solely due to selection effects.  While
Fig.~\ref{f:udrop} clearly shows that we select against low-luminosity
red $U$-dropouts, there is no selection against high luminosity very
blue objects.  The more luminous starbursts have a higher dust content.
Presumably, dust content is related to metallicity.  This is certainly
born out for local starbursts where [O/H] correlates with $\beta$
(Heckman \etal\ 1998). From Paper II, we know that the $U$-dropouts have
similar surface brightnesses, hence the most luminous ones are the
largest, and presumably the most massive galaxies.  Putting these
together implies a mass-metallicity relationship for $U$-dropouts like
that seen in starburst (Heckman \etal\ 1998) and normal galaxies in the
local universe.  The presence of dust in the most luminous $U$-dropouts
implies that they were previously enriched with metals.  Hence we are
not seeing the most luminous galaxies at the instant they ``turn
on''. Their first generation of star formation must have occurred at $z
\geq 2.75$.

\subsection{Optical line to UV continuum ratios as reddening indicators}

Figure~\ref{f:uvem} illustrates the (un) usefulness of line to UV
continuum ratios as indicators of dust absorption; the $y$ axis is
proportional to the ratio of the SFRs one would derive from emission
lines and the UV continuum respectively, in the absence of dust
absorption.  Pettini \etal\ (1998) argue that the similarity of the SFR
derived from \Hbeta\ fluxes and the UV continuum indicates that dust
absorption is not extreme.  However, Fig.~\ref{f:uvem} shows little
correlation between $F_{\rm H\beta}/f_{1600}$ with $\beta$. In local
starburst systems, we know that $\beta$ is a good dust absorption
indicator (Fig.~\ref{f:uvir}); apparently $F_{\rm H\beta}/f_{1600}$ is
not.  The only ratio that noticeably correlates (positively) with
$\beta$ is $F_{\rm H\alpha}/f_{1600}$.  However, the slope of the
correlation is about half that expected by application of the LMC
extinction law.  The $F_{\rm [O\, {\sc iii}]}/f_{1600}$ ratio even shows
an anti-correlation with $\beta$.  This is probably due to the
correlation of [O/H] with $\beta$ found for starbursts (Heckman \etal,
1998) -- highly reddened starbursts have higher metallicities, lower
electron temperatures in the ionized medium, and hence lower [O{\sc
iii}]/H$\beta$ ratios than low reddening starbursts.

The $F_{\rm H\alpha}/f_{1600}$ and $F_{\rm H\beta}/f_{1600}$ results
illustrate a result that is often overlooked: the dust absorption
estimated from continuum observations is usually less than that
estimated from \HII\ emission.  This was first noted by Fanelli, O'Connell,
\&\ Thuan (1988) and subsequently confirmed by others including
Mas-Hesse, Arnault, \&\ Kunth (1989), and Calzetti \etal\ (1994).
The lack of correlation between $F_{\rm H\beta}/f_{1600}$
and $\beta$ indicates that the effective extinction law of
starbursts is such that $A_{1600}({\rm continuum\/}) \approx A_{\rm
H\beta}({\sc Hii\/})$.  While uncorrected UV continuum fluxes and
\Hbeta\ line fluxes may imply the same SFRs, both are
affected by similar amounts of dust absorption.  Hence, emission line
flux to UV continuum ratios do not provide strong constraints on dust
absorption.

\subsection{SCUBA detections and the nature of $U$-dropouts\label{ss:scuba}}

The SFR of the various SCUBA detections is typically derived to be
$> 100\, \Msun\, {\rm yr^{-1}}$ (e.g. Lilly \etal\ 1998).  As
shown in Fig.~\ref{f:lf}, this is consistent with the SFR derived for
the brightest and reddest $U$-dropouts, after dust absorption
correction.  Such high values for the SFR are unprecedented for local UV
selected starbursts, and typically only seen in FIR selected starbursts
such as Arp220.  This suggests that the most luminous starbursts at $z
\approx 3$ are less dust enshrouded than those seen locally - the
ultra-luminous infrared galaxies.  However, this conjecture is difficult
to test, since there are very few UV observations of ultra-luminous
infrared galaxies (cf.\ Trentham \etal\ 1999), hence the amount of UV
absorption they suffer has not been properly characterized.  Also shown
in Fig.~\ref{f:lf} is Lilly's \etal\ (1998) estimate of the number
density of the SCUBA sources in the redshift range $2 < z <3$. This is
also in fair agreement with the number density of the $U$-dropouts
having the same luminosity.  This comparison suggests that the SCUBA
sources may be largely $U$-dropouts with modest amounts of dust
absorption, rather than being completely opaque in the UV.  Furthermore,
it suggests that much of the SFR density at $z \approx 3$ can be
recovered from rest-frame UV measurements alone.

Working against this suggestion is the fact that none of the HDF SCUBA
sources detected by Hughes \etal\ (1998) correspond to known
$U$-dropouts.  However, the numbers are small, and the association with
optical galaxies is still not secure.  Hughes \etal\ estimate $\rho_{\rm
SFR} \approx 0.11\, \Msun\, {\rm yr^{-1}\, Mpc^{-3}}$ at $z \approx 3$
from their SCUBA detections which is about half of the $\rho_{\rm SFR}$
we derive.  This is only $\sim$ five times larger than the M96 estimate
($\sim 3$ times higher than the M98 estimate).  Like us
they do not extrapolate the luminosity function of their detections,
hence the $\rho_{\rm SFR}$ is a lower limit consistent with our
estimates.  Alternatively, if the SCUBA sources are at $z \approx 3$ and
a non-intersecting set with respect to $U$-dropouts (as suggested by the
Hughes \etal\ identifications), then the sum of the Hughes \etal\
$\rho_{\rm SFR}$ and that derived from our $U$-dropout sample provides
an improved lower limit: $\rho_{\rm SFR} \gtrsim 0.30\, \Msun\, {\rm
yr^{-1}\, Mpc^{-3}}$ at $z \approx 3$ which is a factor $\gtrsim 15$
larger than that estimated by M96, and $\gtrsim 9$ times the uncorrected
value estimated by M98.  

M98 argue against such a high value of $\rho_{\rm SFR}$ since it would
over-produce low mass stars in the present universe if it went on for
more than about one Gyr.  However, this argument assumes that
the Lyman-dropouts have an IMF with constant Salpeter (1955) slope over the mass
range of 0.1 to 125 \Msun.  Recent observations of local starbursts
suggest that their IMF is deficient in stars less massive than 3\Msun\
with respect to the Salpeter IMF (Nota \etal\ 1998; Siriani \etal\ 1999)

Since $\rho_{\rm SFR}$ is very model dependent, a more robust prediction
of interest is the lower limit to the FIR emissivity.  Summing the
predicted FIR fluxes of all the HDF $U$-dropouts (using
eq.~\ref{e:irxcal}) and normalizing by the volume yields $\rho_{\rm FIR}
> 7.2\times 10^{29}\, {\rm erg\, s^{-1}\, Hz^{-1} Mpc^{-3}}$ at $z =
2.75$ evaluated at a rest wavelength of 80$\mu$m.  If the SCUBA
detections are too reddened or dimmed to be selected as $U$-dropouts
then we can add them to our sample to get an improved lower limit:
$\rho_{\rm FIR} > 12\times 10^{29}\, {\rm erg\, s^{-1}\, Hz^{-1}
Mpc^{-3}}$. These are still lower limits because completeness
corrections have not been made.  These predictions can be tested
directly with improved FIR, sub-mm, and radio observations (e.g.\ with
the upcoming Sub-mm Array, and the SIRTF, FIRST, and PLANCK space
missions).

\subsection{Application to other redshifts}

Our calibration of UV extinction is readily adapted to observations at
redshifts other than $z \approx 3$ and hence other spectral domains.
The main requirement is that the data include fluxes in two broad bands
or coarse spectra covering the rest frame UV.  Redshift information is
also helpful, and that can be obtained photometrically to adequate accuracy
from complementary rest frame optical - near IR data
(e.g.\ from the Sloan Digital Sky Survey, or deep pointed observations).
For example, serendiptous detections of galaxies with $z \gtrsim 5$
(e.g.\ Franx \etal\ 1997; Dey \etal\ 1998) indicate that galaxy
formation had commenced by this epoch.  One of these galaxies is a
lensed arc at $z = 4.92$, whose broad band (observed frame) optical -
near IR photometry yield $\beta = -1.63$ (Soifer \etal\ 1998), implying
$A_{1600} \approx 1.2$ mag.  Upcoming all sky surveys in the UV by GALEX
(Martin \etal\ 1997) and TAUVEX (Brosch, \etal\ 1994) will probe the
cosmological UV emissivity to $z \lesssim 2$.  Extending deep imaging
with the HST into the near IR with NICMOS allows Lyman edge systems to
be detected out to $z \approx 8$ ($I_{814}$-dropouts).  

There are a few caveats to using our methods, especially at low-$z$.
Firstly, it is based on the assumption that the UV emissivity is
dominated by starbursts.  Possible contaminants are normal disk galaxies
and old stellar populations.  Ideally we want to include the UV light
from normal disk galaxies when estimating star SFRs.  The problem is
that it is not known whether normal disks follow the starburst reddening
law derived here.  Their cool and warm ISM is dominated by a disk
morphology, while starbursts typically have significant quantities of
extra-planar ISM ejected by galactic winds (Heckman, Armus, \&\ Miley
1990).  In Paper I we suggested that winds shape the ISM around
starbursts into a foreground screen, thus producing the \irx\ - $\beta$
correlation.  Further work has to be done to determine if normal disks
have a reddening - UV absorption relation and to calibrate it.
Fortunately normal disk and starburst galaxies can be distinguished in
the UV by effective surface brightness (Paper II).  Cosmological surface
brightness dimming renders normal spiral galaxies difficult to detect by
$z \approx 3$ (Giavalisco \etal\ 1996; Hibbard \&\ Vacca, 1997), hence
they do not contaminate the $U$-dropout sample.  Old stellar populations
produce significant UV ``upturn'' emission.  If they significantly
contaminate a UV survey then the assumption that UV light is due to
recent star formation is seriously compromised.  Old stellar populations
can be distinguished and removed from UV surveys by extending the
broad-band SED into the rest frame optical.

Secondly, our technique has been calibrated to a fairly modest
$A_{1600} \approx 5$ mag (Fig.~\ref{f:uvir}).  While the local
calibrating sample contains several well known dusty starbursts (e.g.\
NGC~1614, NGC~3504, NGC~3256, NGC~7552), it doesn't contain any of the
``Ultra-Luminous Infrared Galaxies'' (ULIRGs) which are the most
luminous starbursts known (Sanders \&\ Mirabel, 1997).  It is not
known if they fall on or continue the \irx\ - $\beta$ relationship
since the number of these galaxies observed in the UV is rather meager
(Trentham \etal\ 1999).  The amount of totally obscured star formation
in the local universe can be estimated by applying our absorption
algorithm to a blind UV survey such as will be produced by GALEX,
after correction for contaminants as suggested above.  This will yield
an estimate of $\rho_{1600,0}$ and also the corresponding density of
UV flux processed to the FIR.  Comparison with $\rho_{\rm FIR}$
derived from IRAS, ISO\footnote{Infrared Space Observatory}, or
SIRTF\footnote{Space Infrared Telescope Facility}
observations will yield the fraction of completely buried star formation.

\section{Summary}\label{s:sum}

The strong similarity between local starburst galaxies and Lyman limit
systems at $z \approx 3$ provides a strong impetus to use starbursts as
templates for interpreting the properties of high-$z$ galaxies.  The
fact that starburst galaxies redden as more of their UV light is
absorbed by dust is demonstrated by a strong correlation between their
FIR to UV flux ratios and their ultraviolet spectral slopes $\beta$
(color).  Our $A_{1600}$ versus $\beta$ calibration is a simple
empirical fit to this correlation.  Hence, by design it recovers the UV
light that has been reprocessed to the FIR.

We applied this calibration to a $U$-dropout sample selected from the
HDF V2.0 catalog.  This sample has an identical limiting mag to that of
Madau \etal\ (1996: M96).  However, the color selection is somewhat
different, resulting in a fairly uniform limiting $A_{1600} \lesssim
3.4$ mag.  In contrast the M96 $U$-dropout sample has a non-uniform bias
against large $A_{1600}$.  For our adopted cosmological parameters of
$H_0 = 50\, {\rm km\, s^{-1}\, Mpc^{-1}}$, $q_0 = 0.5$, the rest-frame
comoving UV emissivity of our sample is $\rho_{1600} = 2.7 \times
10^{26}\, {\rm erg\, cm^{-2}\, s^{-1}\, Hz^{-1}\, Mpc^{-3}}$ without
applying $A_{1600}$ corrections.  This is 1.7 times higher than the
value derived by M96, which illustrates the effects of the differing
color selections.  After correcting for dust absorption we find
$\rho_{1600,0} = 1.4 \times 10^{27}\, {\rm erg\, cm^{-2}\, s^{-1}\,
Mpc^{-3}}$, which is factors of nine and five times higher than the
uncorrected values of M96 and M98, respectively.  However, this
$\rho_{1600,0}$ is still a lower limit because we have not corrected for
incompleteness, and we are not sensitive to galaxies having $A_{1600} >
3.6$ mag.

This study and other recent observations further demonstrate the close
correspondence between local starbursts and high-$z$ Lyman limit
systems.  Firstly, we find a correlation between $\beta$ (color) and
$M_{1600}$ (UV luminosity) for the $U$-dropouts, similar to the
correlation seen in local starburst galaxies (Heckman \etal\ 1998).
To first order the UV luminosity gives the mass of recently formed
stars, while $\beta$ correlates well with metal content in local
galaxies (Heckman \etal\ 1998).  This implies that there is a mass
metallicity relationship out to $z \approx 3$.  Since the highest
luminosity galaxies are clearly reddened they must have produced
enough metals to create the dust prior to the epoch of
observation. Hence the first epoch of star formation occurred at $z
\geq 2.75$ in the most luminous $U$-dropouts.  Secondly, observation
of emission lines in the rest frame optical (observed in the near-IR)
show that $U$-dropouts contain ionizing stellar populations to the
same degree as local starbursts.  This can be seen by comparing the
line flux to UV flux density ratios with $\beta$ for both samples.
However, neither the local nor the $U$-dropout samples show a strong
correlations in these diagrams as would be expected from naive
application of UV extinction/absorption laws.  This suggests that the
column density of dust towards \HII\ emission is higher than towards
the UV continuum both in local starbursts and out to $z \approx 3$.
Consequently, the line flux to UV flux density ratio is not a good
indicator of dust absorption.  Thirdly, the marginal detection of two
HDF $U$-dropouts at radio rest $\lambda_0 \approx 1$ cm by Richards
\etal\ (1998) and non-detections of the $U$-dropouts in the HDF by
Hughes \etal\ (1998) is consistent with estimates of the amount of UV
flux reprocessed into the FIR.  Deeper observations in both domains
(by a factor of $\sim \sqrt{2})$) are required to test the dust
reprocessing model proposed here, and to confirm whether the FIR-radio
correlation holds at $z \approx 3$.

The methodology presented here is readily adapted to other redshifts.
For those sources affected by modest amounts of absorption ($A_{1600}
\lesssim 5$ mag), the intrinsic UV flux can be recovered from UV
quantities alone: flux measurements in two rest frame vacuum-UV bands
are sufficient.  These observations can also be used to estimate the
total UV flux reprocessed into the FIR.  Hence, lower limits to the
intrinsic $\rho_{\rm UV,0}(z)$ and $\rho_{\rm FIR}(z)$ can be
determined.  Integration of $\rho_{\rm FIR}(z)$ and comparison to the
FIR background spectrum (Fixsen \etal\ 1998; Hauser \etal\ 1998) will
constrain the amount of star formation totally obscured by dust from the
ultraviolet.

\acknowledgements We thank two anonymous referees for suggestions that
improved this paper.  Mark Dickinson, Harry Ferguson, Piero Madau, Chris
Martin and Max Pettini stimulated our thinking in numerous discussions.
Piero Madau kindly calculated cosmic transmission curves for us.  We
thank Mark Dickinson for providing the $z$ measurements of the Hawaii and
Berkeley groups collated to the HDF V2.0 catalog.  This research has
made use of the NASA/IPAC Extragalactic Database (NED) which is operated
by the Jet Propulsion Laboratory, California Institute of Technology,
under contract with the National Aeronautics and Space Administration.
This research was supported by NASA grant NAG5-6400.





\end{document}